\documentclass[aps, prd, 
superscriptaddress, nofootinbib, floatfix, noshowpacs]{revtex4-2}

\usepackage{amsmath,amssymb}
\usepackage{color,url}
\usepackage{listings}
\usepackage{slashed}
\usepackage[pdftex]{graphicx}
\usepackage{enumitem}
\usepackage{epstopdf}
\usepackage{epsfig}
\usepackage{grffile}
\usepackage{relsize}
\usepackage{float}
\graphicspath{{./img/}}
\usepackage{soul}

\usepackage{CJKutf8}

\usepackage[mathscr,scaled=1.15]{urwchancal}
\DeclareFontFamily{OT1}{pzc}{}
\DeclareFontShape{OT1}{pzc}{m}{it}%
{<-> s * [1.15] pzcmi7t}{}
\DeclareMathAlphabet{\mathpzc}{OT1}{pzc}{m}{it}

\DeclareMathOperator\atanh{atanh}
\definecolor{purple}{rgb}{0.5,0,0.5}
\definecolor{blue}{rgb}{0.0,0,0.9}
\definecolor{prdblue}{rgb}{0.133,0.118,0.498}
\usepackage[colorlinks=true, pdfstartview=FitV, linkcolor=prdblue, citecolor= prdblue, urlcolor=prdblue]{hyperref}

\RequirePackage{comment}

\begin{document}
\begin{CJK*}{UTF8}{gbsn}

\title{
Pseudoscalar charmonium and bottomonium: light-front wave functions, distribution amplitudes and distribution functions}

\author{Z.-N.\ Xu (徐珍妮)%
    $^{\href{https://orcid.org/0000-0002-9104-9680}{\textcolor[rgb]{0.00,1,0.00}{\sf ID}}}$}
\email[]{zhenni.xu@dci.uhu.es}
\affiliation{Dpto.\ Ciencias Integradas, Centro de Estudios Avanzados en Fis., Mat.\ y Comp., \\
\hspace*{0.5em}Fac.~Ciencias Experimentales, \href{https://ror.org/03a1kt624}{Universidad de Huelva}, E-21071 Huelva, Spain}

\author{Z.-Q.\ Yao (姚照千)%
       $\,^{\href{https://orcid.org/0000-0002-9621-6994}{\textcolor[rgb]{0.00,1,0.00}{\sf ID}}}$}
\email{z.yao@hzdr.de}
\affiliation{\href{https://ror.org/01zy2cs03}{Helmholtz-Zentrum Dresden-Rossendorf}, Bautzner Landstra{\ss}e 400, D-01328 Dresden, Germany}

\author{C. Mezrag%
       $^{\href{https://orcid.org/0000-0001-8678-4085}{\textcolor[rgb]{0.00,1,0.00}{\sf ID}}}$}

\affiliation{Irfu, CEA, Universit\'e Paris-Saclay, 91191, Gif-sur-Yvette, France}

\author{K. Raya%
       $^{\href{https://orcid.org/ 0000-0001-8225-5821}{\textcolor[rgb]{0.00,1,0.00}{\sf ID}}}$}
\affiliation{Dpto.\ Ciencias Integradas, Centro de Estudios Avanzados en Fis., Mat.\ y Comp., \\
\hspace*{0.5em}Fac.~Ciencias Experimentales, \href{https://ror.org/03a1kt624}{Universidad de Huelva}, E-21071 Huelva, Spain}

\author{J.\ Rodr\'iguez-Quintero%
       $^{\href{https://orcid.org/0000-0002-1651-5717}{\textcolor[rgb]{0.00,1,0.00}{\sf ID}}}$}
\affiliation{Dpto.\ Ciencias Integradas, Centro de Estudios Avanzados en Fis., Mat.\ y Comp., \\
\hspace*{0.5em}Fac.~Ciencias Experimentales, \href{https://ror.org/03a1kt624}{Universidad de Huelva}, E-21071 Huelva, Spain}



\begin{abstract}
Light-front wave functions play a central role in the program of understanding the structure of hadrons as QCD bound states. Using continuum Schwinger methods, based on Dyson–Schwinger and Bethe–Salpeter equations, they can be computed directly within a framework connected to QCD. For light pseudoscalar mesons, previous studies revealed an approximate separability of longitudinal and transverse lightcone momentum dependences in the LFWFs, leading to a simple relation between distribution functions  and amplitudes. In this work, we extend those previous studies to the case of pseudoscalar charmonium and bottomonium, using the fictitious $\pi_s$ meson as a benchmark. Motivated by the observed deviations, we propose a modified non-separable LFWF ansatz that successfully reproduces the properties of heavy pseudoscalar quarkonia and allows the calculation of zero-skewness generalised parton distribution functions, electromagnetic and gravitational form factors, and transverse charge and mass distributions.

\end{abstract}
\maketitle
\end{CJK*}

\section{Introduction}

The quantum SU(3) gauge theory, Quantum Chromodynamics (QCD), is our best candidate for describing the internal structure and dynamics of hadrons as the physical bound states of quarks interacting through the mediation of gluons. Our understanding of this internal structure based on QCD will be challenged in the next few years by the future Electron Ion Collider\,\cite{Accardi:2012qut,Arrington:2021biu}, and by the collaborative and complementary effort at other modern facilities\,\cite{Anderle:2021wcy,Quintans:2022utc,Accardi:2023chb}. Quantities such as generalized parton distributions (GPDs)\,\cite{Dittes:1988xz,Muller:1994ses,Radyushkin:1996nd,Ji:1996nm} or transverse momentum distributions (TMDs)\,\cite{Pasquini:2014ppa}, which reveal key properties of hadrons and enable a spatial picture of their charge and mass distributions\cite{Polyakov:2018zvc,Diehl:2003ny,Burkardt:2000za,Burkardt:2002hr}, will be probed with unprecedented precision and need therefore to be computed as accurately as possible within a framework directly connected to QCD.   

These distributions are deeply underlain by the light-front wave functions (LFWFs)\,\cite{Brodsky:1979gy}, the components of the expansion of a given hadron in Fock space and the quantum-field-theory analogue of quantum-mechanics wave functions, providing a probabilistic interpretation for the distribution functions, especially for those emerging in the so-called forward limit of the GPDs: the parton distribution functions (PDFs). LFWFs also remain a basic building block for a practical bridge from PDFs to the distribution amplitudes (DAs), emerging from GPDs in a different kinematical limit\footnote{Particularly, in the chiral limit, soft-pion theorems connects the maximally skewed pseudoscalar GPDs to DAs\,\cite{Polyakov:1998ze}.}, which has very often been used in recent literature in the context of the calculations of meson distribution functions and amplitudes based on the solutions of Dyson-Schwinger (DSEs) and Bethe-Salpeter equations (BSEs)\,\cite{Chang:2013pq,Chang:2014lva,Cui:2020tdf,Cui:2020dlm,Yao:2025xjx,Arrington:2021biu,Raya:2022eqa,Raya:2024ejx,Roberts:2021nhw,Yao:2025xjx}, as well as on the perturbation-theory integral representation (PTIR) of the solutions for BSEs\,\cite{Chang:2013pq,Chang:2014lva,Mezrag:2014jka,Chouika:2017rzs,Zhang:2021mtn,Roberts:2021nhw,Raya:2021zrz,Chavez:2021llq,Chavez:2021koz,Albino:2022gzs,Almeida-Zamora:2023rwg,Almeida-Zamora:2023bqb}. Indeed, the LFWFs can be directly computed as the appropriate light-front projections of the solutions for the BSEs\,\cite{Chang:2013pq}, namely the bound-state solutions for the two-body problem in relativistic quantum field theory. The projections also involve the quark propagators, which are the Schwinger functions obtained as solutions of the gap equation, the two-body quark DSE. Remarkably, after a sensible choice of the interaction kernels in the DSEs and BSEs, one is left with solutions entailing LFWFs from which mesons distribution functions and amplitudes are obtained, thereby yielding a computational framework directly connected to QCD, that is commonly referred to in literature as continuum Schwinger methods (CSM)\,\cite{Roberts:2012svX,Binosi:2014aea,Eichmann:2016yit,Qin:2020rad}. 

Importantly, it should be underlined that, within this CSM framework, the outputs are to be interpreted as resolved at the hadron scale, $\zeta_H$, at which all hadron structural properties can be expressed in terms of quasiparticle valence degrees of freedom, namely the fully dressed constituent quarks at such scale. Therefore, the PDF in this framework must be understood as a quasiparticle distribution function, expressing the probability for the valence quasiparticle to carry a given lightcone momentum fraction in a hadron. QCD all-orders evolution\cite{Cui:2020tdf,Raya:2021zrz,Cui:2021mom,Cui:2022bxn,Lu:2022cjx,Yin:2023dbw} can then be implemented to make contact with the empirical world at any larger scale, $\zeta > \zeta_H$. However, scale-independent quantities such as electromagnetic and gravitational form factors, as well as the corresponding charge and mass distributions and radii, can be directly evaluated at the hadron scale without evolution (see, \emph{e.g.}\,, Refs.\,\cite{Raya:2021zrz,Raya:2024ejx,Raya:2024glv,Miramontes:2025vzb,Miramontes:2026san,Broniowski:2024oyk,Ydrefors:2021dwa,Xu:2023bwv}).                 

In the last decade, a plethora of works capitalised on the previous computational framework, most especially for pseudoscalar mesons, delivering a large body of evidence connecting their properties and the emergence of hadron mass (EHM); \emph{e.g.}, see Refs.\,\cite{Roberts:2016vyn,Roberts:2021nhw,Arrington:2021biu,Anderle:2021wcy,Quintans:2022utc,Accardi:2023chb}. The special case of the pion is of particular importance, owing to its double role as the lightest strongly interacting particle --and thereby an effective mediator of the nuclear forces-- and as a near Nambu-Goldstone boson, where only the small masses of the constituent quark and antiquark break explicitly the chiral symmetry (\emph{e.g}, see Ref.\,\cite{Horn:2016rip}). The interplay between effects from explicit and dynamical chiral symmetry breaking in EHM is driven by the mass of these constituents, and this motivates also notable interest in the study of other light pseudoscalars as the kaon or the fictitious system made of two first-generation quark and antiquark constituents with degenerate strange mass, $\pi_s$\,\cite{Xu:2018eii,Chen:2016sno,Cui:2020tdf,Cui:2020dlm,Zhang:2021mtn,Raya:2021zrz,Xing:2025eip,Yao:2025xjx,Cheng:2026nud}. For all of them, their LFWFs are found to be endowed with an approximate property which has been extensively exploited: the dependence on the lightcone longitudinal and transverse momenta decouples, entailing a clean and direct connection between the corresponding DAs and PDFs, \emph{i.e.}, the latter being proportional to the square of the former. This latter property, that the LFWF is separable, paves the way to accessing GPDs, thereby deriving important implications for meson internal structure, from the knowledge of PDFs supplemented with little phenomenological information\,\cite{Raya:2021zrz,Xu:2023bwv,Xu:2024nzp,Raya:2024ejx}.    

Recently, DAs and PDFs for the pseudoscalar charmonium have been calculated within the same CSM framework described above\,\cite{Zeng:2026peb}. Their scrutiny reveals that the behavior of the PDF slightly deviates from that of the squared DA, implying that a separable LFWF cannot fully express the  structural properties of the pseudoscalar charmonium. In the current work, we calculate DAs and PDFs for both pseudoscalar charmonium and bottomonium, considering also the $\pi_s$ as a benchmarking case. Confirming the deviations when comparing PDFs and squared DAs for the heavy quarkonia, we also reproduce here the analysis based on a direct, symmetry-preserving calculation of LFWFs, previously used in Ref.\,\cite{Yao:2025xjx} to demonstrate the validity of a separable LFWF for the pion and $\pi_s$. Building on the results of this latter analysis, we propose to amend the LFWF separable behavior by the introduction of a function of the lightcone momentum that modifies the mass dimension, thereby accounting successfully for the results obtained for the pseudoscalar quarkonia LFWFs, DAs and PDFs. 

The presentation of the manuscript is arranged as follows. We introduce some generalities about LFWFs, DAs and PDFs in section \ref{sec:LFWF}, while the prescription for a non-separable LFWF is presented and discussed in section \ref{sec:nonsep}. The CSM calculations and the practical validation of this prescription can be found in section \ref{sec:DSEresults}. Section \ref{sec:illustration} is then devoted to illustrate that the prescription can be used to extend a pion LFWF model (appendix \ref{app:GPD}), used in the literature to calculate the beam spin asymmetry in deeply virtual pion Compton scattering (DVCS)\,\cite{Chavez:2021koz,Chavez:2021llq}, thus delivering the zero-skewness GPDs for heavier pseudoscalar quarkonia, and the corresponding electromagnetic and gravitational form factors, together with the charge and mass distributions in the lightcone transverse plane. Finally, we conclude in section \ref{sec:conclusions}.

\section{LFWFs and distribution functions}
\label{sec:LFWF}

Let us consider the pseudoscalar meson-$\mathsf{P}$ LFWF\,\cite{Yao:2025xjx}, 
\begin{equation}
\psi_{\mathsf P}(x,k_\perp^2)
 \propto
\gamma_5 \big[\gamma\cdot n \,\psi_{\mathsf P}^{0}(x,k_\perp^2)\nonumber + i \sigma_{\mu\nu}n_\mu k_{\perp \nu} \, \psi_{\mathsf P}^{1}(x,k_\perp^2)\big]\,,
\label{EqLFWF}    
\end{equation}
where $n$ is a lightlike $4$-vector, $n^2=0$, $n\cdot k_\perp = 0$, and $n\cdot P=-m_{\mathsf 5}$ in the meson rest frame. It contains the two components labelled by the superscript on $\psi$, indicating the light-front orbital angular momentum projection: the ${\mathpzc L}=0=\uparrow\downarrow=\downarrow\uparrow$ wave function has the light-front spins of the valence constituents antialigned; and ${\mathpzc L}=1=\uparrow\uparrow=\downarrow\downarrow$ has them aligned. They can be obtained via light-front projections of the Bethe-Salpeter wave function (BSWF), ${\mathpzc X}_{\mathsf P}(k;P)$, as follows\,\cite{Chang:2013pq}:   
\begin{equation}
    \psi_{\mathsf P}^{\mathpzc L}(x,k_\perp^2)
     = {\rm tr}_{\rm CD}\int\frac{dk_3 dk_4}{\pi}
    \delta(x n\cdot P - n\cdot k_\eta)
    {\mathpzc P}^{\mathpzc L} {\mathpzc X}_{\mathsf P}(k;P)\,,
    \label{LFWFP}
\end{equation}
with $k$ for the relative momentum between the valence quark and antiquark, and $k_\eta = k+\eta P$, $0\leq \eta \leq 1$ (for mass-degenerate valence quark and antiquark, it is useful to choose $\eta=1/2$); the trace is over colour and spinor indices
and
\begin{equation}
{\mathpzc P}^{0} = \frac{1}{4}\gamma_5\gamma\cdot n \,,
\quad {\mathpzc P}^{1} = \frac{1}{4}\frac{i}{k_\perp^2} \gamma_5 \sigma_{\mu\nu} n_\mu k_{\perp\nu}\,.
\end{equation}
In the next section, the BSWFs ${\mathpzc X}_{\mathsf P}(k;P)$ will be obtained by solving the corresponding Bethe-Salpeter and Dyson-Schwinger equations and, in the aim of circumventing the need of more detailed knowledge of ${\mathpzc X}_{\mathsf P}(k;P)$ in the complex plane, we shall proceed by the calculation of the following $k_\perp^2$-dependent Mellin moments:
\begin{align}
\label{eq:moments}
\langle x^m  \rangle^{\mathpzc L} (k_\perp^2)
 = \int_0^1 dx\,x^m \, \psi_{\mathsf P}^{\mathpzc L}(x,k_\perp^2) 
= \frac{1}{n\cdot P }{\rm tr}_{\rm CD}\int\frac{dk_3 dk_4}{\pi}
\, \left[\frac{n\cdot k_\eta}{n\cdot P}\right]^m
{\mathpzc P}^{\mathpzc L} {\mathpzc X}_{\mathsf P}(k;P)\,;
\end{align}
by using the approach discussed in connection with Ref.\,\cite[Eq.\,(15)]{Xu:2025cyj}; and obtaining stable results for the $k_\perp^2$-dependence of moments $m=0,1,\ldots,7$ for pseudoscalar charmonium and bottomonium states, \emph{i.e.}, $\mathsf{P}=\eta_c,\eta_b$.

Then, after reconstructing a LFWF from these sets of moments by applying the physically constrained procedure described in Refs.\,\cite{Chang:2013pq, Cloet:2013tta, Cui:2020tdf,Cui:2022bxn, Xu:2024nzp}, the leading-twist two quasiparticles DA and PDF can be straightforwardly obtained from Eq.\,\eqref{LFWFP} as\,\footnote{Since we are dealing with ``isospin''-symmetric states, there is no need to distinguish between particle and antiparticle valence constituents.} 
\begin{subequations}
\begin{eqnarray}
f_{\mathsf P} \varphi_{\mathsf P}(x) &=&
\int \frac{d^2k_\perp}{ 16\pi^3}
 \psi_{\mathsf P}^{0}(x,k_\perp^2)\,,
 \label{DefineDA} \\
{\mathpzc q}_{\mathsf P}(x) &=& 
\int 
\frac{d^2k_\perp}{16\pi^3} \, 
\left[ | \psi_{\mathsf P}^{0}(x,k_\perp^2)|^2
+ k_\perp^2 | \psi_{\mathsf P}^{1}(x,k_\perp^2)|^2\right] \,,
\label{DefPDF}
\end{eqnarray}
\end{subequations}
where $f_{\mathsf P}$ is the pseudoscalar meson-$\mathsf{P}$'s leptonic decay constant. Alternatively, combining Eqs.\,\eqref{eq:moments} and \eqref{DefineDA}, the Mellin moments of the DA can be directly evaluated as
\begin{align}\label{eq:DAfromBSA}
f_\mathsf{P} \langle x^m \rangle_{\varphi_\mathsf{P}} = f_\mathsf{P} \int_0^1 dx \, x^m \varphi_\mathsf{P}(x) = \int \frac{d^4k}{(2\pi)^4} 
\, \left[\frac{n\cdot k_\eta}{n\cdot P}\right]^m
{\mathpzc P}^{0} {\mathpzc X}_{\mathsf P}(k;P)\,,
\end{align} 
where a translationally invariant regularisation scheme should be considered to calculate the four-dimensional integral (the same is also implicit for Eq.\,\eqref{DefineDA}); and these moments can be further used for the reconstruction of the distribution. 

\section{A non-separable LFWF}
\label{sec:nonsep}

Following the analysis of the $k_\perp^2$-dependent Mellin moments for $\mathsf{P}=\pi,\pi_s$ from Ref.\,\cite{Yao:2025xjx}, it can be concluded that both LFWF spin components can very approximately be written in a factorised form:
\begin{equation}
\label{EqSep}
\psi_{\pi,\pi_s}^{\mathpzc L}(x,k_\perp^2)
\approx \varphi_{\pi,\pi_s}(x) \times
F^{\mathpzc L}(k_\perp^2)\,;
\end{equation}
which is manifestly consistent with Eq.\,\eqref{DefineDA} and, more importantly, entailing from Eq.\,\eqref{DefPDF} that
\begin{equation}\label{eq:PDFeqDA2}
    {\mathpzc q}_{\pi,\pi_s}(x) \propto \varphi_{\pi,\pi_s}^2(x)\,;
\end{equation}
while analyses in perturbative QCD predict that \cite{Lepage:1980fj}  
\begin{equation}
    k^2 \gg m_N^2 \, | \quad
    F^{0}(k_\perp^2) \propto \frac{1}{k_\perp^2}
    \propto k_\perp^2 F^{1}(k_\perp^2)\,,
    \label{LFWFUV}
\end{equation}
up to (damping) $\ln k_\perp^2$ corrections, instructing us to introduce a particular suitable form for $F^{\mathpzc L}(k_\perp^2)$ in Ref.\,\cite[Eqs.\,(17,19)]{Yao:2025xjx}. As bound-states systems made of up, down and strange quarks possess LFWFs obeying the separability given by Eq.\,\eqref{EqSep}, the same is expected for the kaon and, in practice, the analogue of Eq.\,\eqref{eq:PDFeqDA2} has been applied for the kaon in Ref.\,\cite{Cui:2020dlm}.    

This factorised form may be expected to be flawed by the increasing mass of the pseudoscalar bound state, which is augmented by the Higgs-induced explicit symmetry breaking impacting on the dressed quark mass. This is confirmed by extending to heavier quarkonia the very same analysis performed in Ref.\,\cite{Yao:2025xjx}, by scrutinising the ratios:
\begin{align}
\label{eq:ratios}
R_\mathsf{P}^{m,\, \mathpzc{L}}(k_\perp^2) =
\frac{\langle x^m \rangle_\mathsf{P}^{\mathpzc L}(k_\perp^2)}{\langle x^0 \rangle_\mathsf{P}^{\mathpzc L}(k_\perp^2)}; 
\end{align}
obtained from Eq.\,\eqref{eq:moments}. They, in the cases of pseudoscalar charmonium and bottomonium displayed in Fig.\,\ref{fig:DFvsDA2heavy}, contrarily to the pion and $\pi_s$\cite[Fig.\,2]{Yao:2025xjx}, show a clear, well-defined pattern, which differs from a constant $k_\perp^2$-independent profile and exposes a small but apparent breakdown of the factorisation represented by Eq.\,\eqref{EqSep}. This observation is consistent with the conclusions presented in Ref.\,\cite{Zeng:2026peb}, where small deviations with respect to an analogue of Eq.\,\eqref{eq:PDFeqDA2} were reported in an   analysis of the charmonium distribution function and amplitude.

A simple and very general prescription to account for these small violations of the factorisation, instructed by the discussions in Ref.\,\cite{Raya:2024glv} and the analysis of Ref.\,\cite{Albino:2022gzs} based on a perturbation theory integral representation (PTIR) of the Bethe-Salpeter amplitude (BSA), can be implemented as follows. 

First, a dimensional analysis of Eq.\,\eqref{DefineDA} tells us that $[F^{0}(k_\perp^2)/f_\mathsf{P}] = 1/M^2$, with $M$ denoting a dimension of mass. Then, considering that any mass dimension other than the leptonic decay constant can be referred to the dressed quark mass, $M_q$, when representing the LFWF, one should expect that: 
$F^0(k_\perp^2) \equiv \overline{F}^0(k_\perp^2/M_q^2) f_\mathsf{P}/M_q^2$. Thus, we can introduce the following dimensionless quantity, $k_\perp^2 = M_q^2 \,t$, and, working with Eqs.\,\eqref{DefineDA} and \eqref{EqSep}, obtain 
\begin{align}
\int \frac{d^2 k_\perp^2}{16 \pi^3} \, \psi_{\mathsf P}^{0}(x,k_\perp^2) = f_\mathsf{P} \, \varphi_\mathsf{P}(x) \int_0^\infty \frac{dt}{16\pi^2} \overline{F}^0(t) \;,  
\label{DefineDA2}
\end{align}
which fixes the normalisation of $\overline{F}^0$ according to Eq.\,\eqref{DefineDA}. In the case of $F^1$, the dimensional homogeneity of Eq.\,\eqref{DefPDF} --consistent with Eq.\,\eqref{LFWFUV}-- and the fact that both LFWF components derive from light-front projections of the BSWF $\chi_\mathsf{P}$ suggest that\footnote{This is illustrated by, \emph{e.g.}, \cite[Eqs.\,(21a,21b)]{Raya:2026gwg}, where both LFWF components, and hence $F^0$ and $F^1$, were obtained from the light-front projections of a PTIR of the BSA.
} $F^1(k_\perp^2) \equiv \overline{F}^1(k_\perp^2/M_q^2) f_\mathsf{P}/M_q^3$.

Then, a minimal modification of Eq.\,\eqref{EqSep}, introducing an explicit breakdown of the factorisation but letting Eq.\,\eqref{DefineDA} trivially unchanged, can be obtained by replacing both in $F^0$ and $F^1$, 
\begin{equation}\label{eq:modif}
M_q^2 \rightarrow M_q^2 \kappa(x) \;,    
\end{equation}
where $\kappa(x)$ is a given function of $x$ that will be characterised later below. It can be immediately seen that Eq.\,\eqref{DefineDA2}, and hence Eq.\,\eqref{DefineDA}, remain both unaltered by the replacement \eqref{eq:modif}. Furthermore, reminding that Eq.\,\eqref{DefPDF} implies Eq.\,\eqref{eq:PDFeqDA2} for a separable LFWF, one can straightforwardly prove that, after applying \eqref{eq:modif}, Eq.\,\eqref{DefPDF} entails 
\begin{align}\label{eq:q5vsDA2}
{\mathpzc q}_\mathsf{P}(x) 
= f^2_\mathsf{P} \frac{\varphi_\mathsf{P}^2(x)}{M_q^2 \kappa(x)}
\, \int_0^\infty \frac{dt}{16\pi^2} \left( \left[\overline{F}^0(t)\right]^2 + t \left[\overline{F}^1(t)\right]^2 \right)
\propto \frac{\varphi_\mathsf{P}^2(x)}{\kappa(x)} \;, 
\end{align}
thus extending Eq.\,\eqref{eq:PDFeqDA2} to the case of heavier quarkonia. On the other hand, working with Eqs.\,\eqref{eq:moments}, applying the same minimal modification of Eq.\,\eqref{EqSep} for a non-separable LFWF and following the same dimensional argument exposed above, one can conclude that 
\begin{align}\label{momentskappa}
\langle x^m \rangle_\mathsf{P}^0(0) = \overline{F}_0(0)
\frac{f_\mathsf{P}}{M_q^2} \int_0^1 dx \, x^m \frac{\varphi_\mathsf{P}(x)}{\kappa(x)} \;.
\end{align}
While, stemming from it and from Eq.\,\eqref{eq:ratios}, we obtain 
\begin{subequations}
\label{eq:momentstilda}
\begin{align}
\label{eq:R0}
&R_\mathsf{P}^{m, 0}(0) = \langle x^m \rangle_{\widetilde{\varphi}_{\mathsf P}} = \int_0^1 dx \, x^m \widetilde{\varphi}_{\mathsf P}(x) \;, \\ 
\label{eq:varphitilda}
\mathrm{with} \quad&\widetilde{\varphi}_{\mathsf P}(x) = N_\kappa \, \frac{\varphi_{\mathsf P}(x)}{\kappa(x)} \;, \quad 
N_\kappa^{-1} = \int_0^1 dx \, \frac{\varphi_{\mathsf P}(x)}{\kappa(x)} \;.
\end{align}     
\end{subequations}
Here, $\widetilde{\varphi}_\mathsf{P}(x)$ behaves as a new normalised distribution which differs from  $\varphi_{\mathsf P}(x)$ by the factor $1/\kappa(x)$, as ${\mathpzc q}_{\mathsf P}(x)$ from $\varphi_\mathsf{P}^2(x)$ according to Eq.\,\eqref{eq:q5vsDA2}. 

As can be seen in the right panels of Fig.\,\ref{fig:DFvsDA2heavy}, the comparison of the pseudoscalar quarkonia squared DAs and PDFs, both obtained by solving the corresponding Bethe-Salpeter and Dyson-Schwinger equations (see Sec.\,\ref{sec:DSEresults}), is very instructive: the effect of the factor $1/\kappa(x)$ is compressing the distribution on $x$, while respecting the symmetry under $x \leftrightarrow 1-x$. Furthermore, respecting the PDF end-points behavior also entails $\kappa(x) \to 1$ both when $x \to 0$ and $x \to 1$. 

Then, Eq.\,\eqref{eq:R0} indeed explains the pattern shown by the right panels of Fig.\,\ref{fig:DFvsDA2heavy}.  As discussed in Ref.\,\cite{Cui:2022bxn}, for a distribution of support $x \in [0,1]$, symmetric around $x=1/2$, the more compressed it is the lower their Mellin moments are, such that 
\begin{align}\label{eq:constraints}
\frac{1}{2^m} <  \langle x^m \rangle_{\widetilde{\varphi}_{\mathsf P}} < \langle x^m \rangle_{\varphi_{\mathsf P}} < \frac{1}{1+m}   \;,
\end{align}
for $m > 1$ (in the case $m=1$: $\langle x^m \rangle_{\widetilde{\varphi}_{\mathsf P}} = \langle x^m \rangle_{\varphi_{\mathsf P}} = 1/2$). On the other hand,    
working with Eqs.\,\eqref{eq:moments},  \eqref{DefineDA} and \eqref{eq:ratios}, it can be straightforwardly proved that 
\begin{equation}\label{eq:intR0}
\int d^2k_\perp \, R_\mathsf{P}^{0,m}(k_\perp^2) = \langle x^m \rangle_{\varphi_{\mathsf P}} \,. 
\end{equation}
This is precisely what the right panels of Fig.\,\ref{fig:DFvsDA2heavy} show, \emph{i.e.}, for all $m$, the curves for the ratios $R_\mathsf{P}^{m, 0}(k_\perp^2)$ lie systematically below, at $k_\perp^2=0$, the horizontal lines representing the DA Mellin moments of order $m$, as imposed by \eqref{eq:constraints}, then cross them at a given point and remain above them at larger momenta. This crossing is required to satisfy Eq.\,\eqref{eq:intR0}.    

More interestingly, a simple and effective quantitative analysis can be constructed from the ratios evaluated at $k_\perp^2=0$. To this end, the function $\kappa(x)$ introduced in Eq.\,\eqref{eq:modif} and appearing in Eqs.\,(\ref{eq:q5vsDA2}–\ref{eq:momentstilda}) is to be modeled by a given parameter-dependent form $\kappa_\alpha(x)$, with a parameter $\alpha$ determined by minimising
\begin{equation}\label{eq:chi2}
\chi^2(\alpha)= \sum_{m=2}^N \left(R_\mathsf{P}^{0,m}(0) - N_{\kappa_\alpha} \int_0^1 dx x^m \frac{\varphi_\mathsf{P}(x)}{\kappa_\alpha(x)}\right)^2 \;, 
\end{equation}
where the ratios $R_\mathsf{P}^{0,m}(0)$ and the DA $\varphi_\mathsf{P}(x)$ are either calculated or reconstructed, respectively, with the Mellin moments \eqref{eq:moments} and \eqref{eq:DAfromBSA}, obtained from the BSWFs. Then, once $\kappa_\alpha(x)$ determined and identified with $\kappa(x)$, the quasiparticle PDF $\mathpzc{q}_\mathsf{P}$ is given by Eq.\,\eqref{eq:q5vsDA2}, obtained from nothing but the DA and the Mellin moments of the LFWF at zero momentum. 

\section{Results from Continuum Schwinger Methods}
\label{sec:DSEresults}

\begin{figure}[t]
    \centering
    \begin{tabular}{cc}         
            \includegraphics[width=0.43\textwidth]{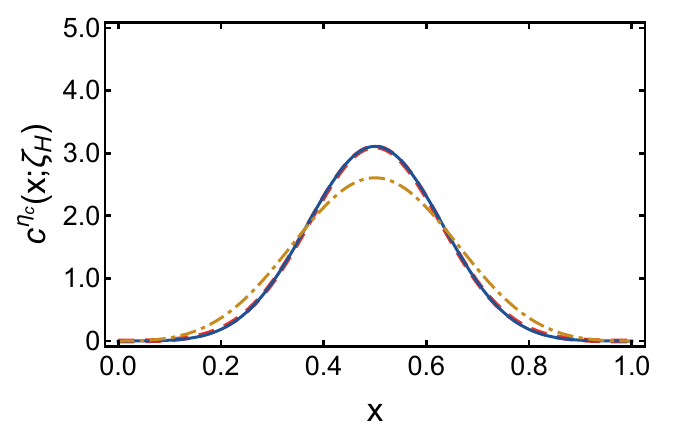}   &
            \hspace{-0.12\textwidth}
            \includegraphics[width=0.75\textwidth]{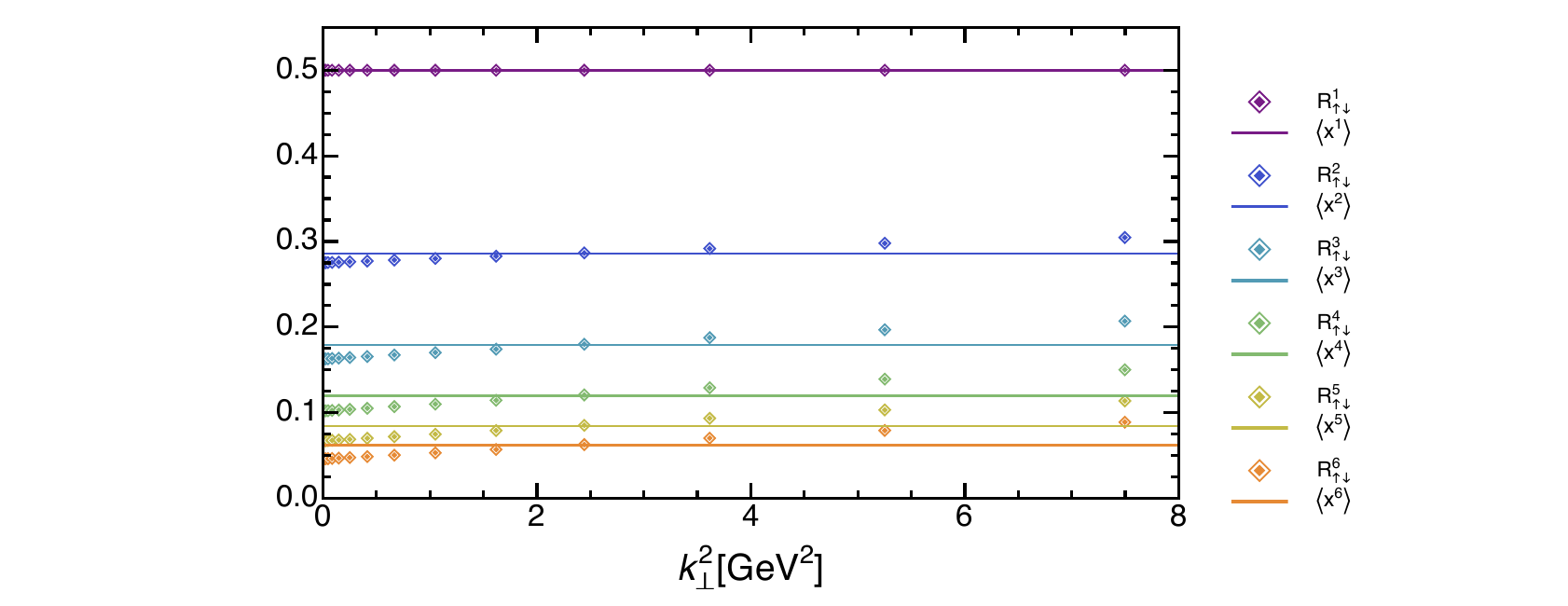} \\ 
            \includegraphics[width=0.43\textwidth]{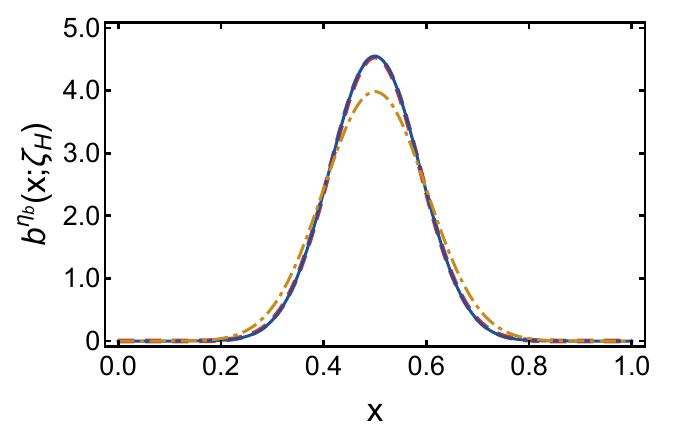}   & 
            \hspace{-0.12\textwidth}
            \includegraphics[width=0.75\textwidth]{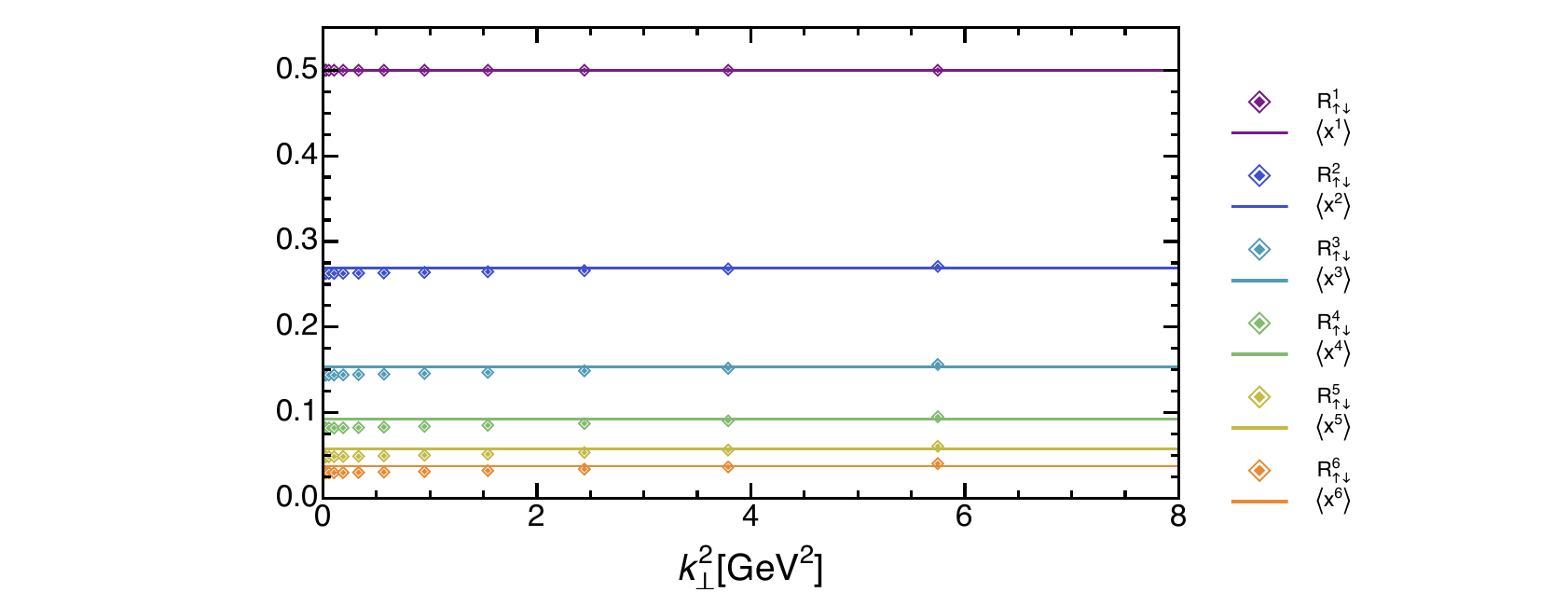}
    \end{tabular}
    \caption{
Results for $\eta_c$ (upper row) and $\eta_b$ (lower row) at the hadron scale $\zeta_H$.
Left panels: quasiparticle distribution functions obtained directly from DSEs (blue solid curves), from the LFWF construction (red dashed curves), and the normalized squared DAs (golden dash-dotted curves).
Right panels: ratios of the $k_\perp^2$-dependent LFWF Mellin moments,
$R_P^{m,L}(k_\perp^2)$, compared with the corresponding DA Mellin moments $\langle x^m\rangle$ shown as horizontal lines.
}
    \label{fig:DFvsDA2heavy}
\end{figure}

\begin{table}[!h]
\caption{Mesons' masses and decay constants for pion and pseudoscalar quarkonia herein under study, obtained with the RL quark + quark scattering kernel described with Eqs.\,(\ref{eq:kernelRL}-\ref{eq:GQC}), compared to their experimental counterparts. Although not a focus of the present analysis, kaon mass and decay constant are also seen to compare excellently well with the experiment: 0.494 and 0.110 GeV (herein and for the experimental central values).}
\label{tab:spectrum}
\centering
\begin{center}
\begin{tabular}{cccccccccc}
 $\textrm{P}$ & $\pi$ & $\eta_s$ & $\eta_c$ & $\eta_b$ \\
 \hline
 \hline
$m_\textrm{P}$ (Herein)   &0.14& 0.69 & 2.98 & 9.4 \\
\hline
$m_\textrm{P}$ (exp.)   &0.14 & - & 2.98 & 9.4 \\
\hline
 \hline
$f_\textrm{P}$ (Herein) &0.093 &0.134 & 0.284 & 0.565 \\
\hline
$f_\textrm{P}$ (exp.) & 0.093(1)& - & 0.237(52) & - \\
\hline
\end{tabular}
\end{center}
\end{table}

Our starting point in this section is obtaining the LFWF and DA moments for pseudoscalar charmonium and bottomonium states, by evaluating, respectively, \eqref{eq:moments} and \eqref{eq:DAfromBSA}. For the sake of comparison, we have also considered the case of $\eta_s$, a fictitious neutral pseudoscalar bound state of two light quarks with strange current mass, which shares the same DA and PDF as the also fictitious $\pi_s$.   
To the goal of obtaining the moments, one must first solve the corresponding Bethe-Salpeter and gap equations, from which the BSWF ${\mathpzc X}_{\mathsf P}(k;P)$ entering in both Eqs.\,(\ref{eq:moments}-\ref{eq:DAfromBSA}) are calculated. The central element to achieve these solutions is the scattering kernel, which, in the Rainbow-Ladder (RL) truncation, takes the form\,\cite{Maris:1997tm}:
\begin{align}\label{eq:kernelRL}
\mathcal{K}^{rs}_{tu}(k) = \mathcal{G}(k^2) T_{\mu\nu}(k) 
\left[ i \gamma_\mu \frac{\lambda^a} 2 \right]_{st} 
\left[ i \gamma_\nu \frac{\lambda^a} 2 \right]_{ur} \;, 
\end{align}
with $T_{\mu\nu}(k)=k^2 \delta_{\mu\nu} - k_\mu k_\nu$, specifying the usual Landau gauge, and $a,r,s,t,u$ representing colour and spinor indices.

The quark-gluon interaction in the kernel is effectively expressed in this work by\,\cite{Qin:2011dd,Binosi:2014aea}  
\begin{align}\label{eq:GQC}
\mathcal{G}(y)=\frac{8\pi^2}{\omega^4} D e^{-y/\omega^2} + 
\frac{8\pi^2 \gamma_m \mathcal{F}(y)}{\ln\left[ \tau + (1+y/\Lambda_\textrm{QCD}^2)^2\right]} \;,
\end{align}
with $\gamma_m$=12/25, $\Lambda_\textrm{QCD}$=0.234 GeV, $\tau=e^2-1$ and $\mathcal{F}(y)=1-\exp{(-y/\Lambda_I^2)}$, $\Lambda_I=1$ GeV; while the parameters $D$ and $\omega$ used in this work are phenomenologically set to $\omega$=0.5 GeV, $D\omega$=$(0.80\,\mathrm{GeV})^3$ in the light sector and $\omega$=0.8 GeV, $D\omega$=$(0.47\,\mathrm{GeV})^3$ in the heavy, such that meson's masses and decay constants are obtained in agreement with the experiment (see Tab.\,\ref{tab:spectrum}). 

Once the DA reconstructed from its moments, the PDF can be also obtained working with Eq.\,\eqref{eq:q5vsDA2} and, following the procedure described in the previous section around Eq.\,\eqref{eq:chi2}, with the LFWF moments. This last calculation only relies on the very general assumption that a separable LFWF, proved to work fairly well for light pseudoscalar mesons, can be extended to the heavier charmonium and bottomonium pseudoscalar states only implementing the minimal modification \eqref{eq:modif}. This calculation and its relying assumption for the LFWF can be tested because, equivalently, the PDF can be also directly calculated from the BSWF as\,\cite{Chang:2014lva,Mezrag:2014jka,Ding:2019lwe,Ding:2019qlr}
\begin{align}\label{eq:DSEPDF}
{\mathpzc q}_\mathsf{P}(x;\zeta) = \textrm{tr}_\textrm{CD} \int_{dk}
\delta^x_{n,P}(k_\eta) 
\Gamma_\mathsf{P}(k_{\bar{\eta}\eta},P;\zeta) S(k_{\bar{\eta}};\zeta)\, 
n \cdot \partial_{k_\eta}\left\{ \Gamma_\mathsf{P}(k_{\eta\bar{\eta}},-P;\zeta) S(k_\eta;\zeta)\right\} \;, 
\nonumber 
\end{align}
where  
\begin{align}
\chi_\mathsf{P}(k_{\eta\bar{\eta}},P;\zeta) = S(k_\eta;\zeta) \Gamma_\mathsf{P}(k_{\eta\bar{\eta}},P;\zeta) 
S(k_{\bar{\eta}};\zeta) \;,
\end{align}
with $S(k;\zeta)$ representing the involved quark propagator and $k_\eta=k+\eta P$, $k_{\bar{\eta}}=k-(1-\eta)P$ and $k_{\eta\bar{\eta}}=[k_\eta+k_{\bar{\eta}}]/2$, the PDF becoming independent of any $\eta \in [0,1]$. Here, on the one hand, it should be noted that we have chosen to make explicit the momentum scale $\zeta$ at which the distribution function is obtained from Eq.\,\eqref{eq:DSEPDF}. On the other hand, following Ref.\,\cite{Cui:2020tdf}, two special features of Eq.\,\eqref{eq:DSEPDF} should be highlighted: 
\begin{itemize}
\item (i) when evaluated within a rainbow-ladder truncation approach, it can be straightforwardly proved from multiplicative renormalization that $\Gamma_{\mathsf{P}} S$ appears to be a renormalization scale invariant and, consequently, the PDF $\mathpzc{q}_\mathsf{P}$ thus computed happens also to remain scale invariant; 
\item and (ii) that, resulting from Eq.\,\eqref{eq:DSEPDF}, $\mathpzc{q}_\mathsf{P}(x;\zeta)=\bar{\mathpzc{q}}_\mathsf{P}(1-x;\zeta)$ and, hence, $\langle x \rangle_{{\mathpzc q}_\mathsf{P}}^\zeta + \langle x \rangle_{\bar{\mathpzc q}_\mathsf{P}}^\zeta=1$, for quark and anti-quark momentum fractions from the PDFs obtained with \eqref{eq:DSEPDF} for a $q\bar{q}$ bound-state. 
\end{itemize}
The main implication of these two features is that Eq.\,\eqref{eq:DSEPDF} indeed exposes structure properties of a meson at the resolving hadron scale, $\zeta=\zeta_H$, at which the total meson's momentum is shared by the valence-quark and -antiquark, to be then interpreted as fully dressed quasiparticles and the bound-state's degrees of freedom expressing all measurable properties of the meson\,\cite{Ding:2019lwe,Ding:2019qlr,Cui:2020tdf,Raya:2024ejx}. Implicitly, this is the momentum scale considered for the DAs in the previous sections, and the scale at which Eqs.\,\eqref{eq:PDFeqDA2} and \eqref{eq:q5vsDA2} are reliable, respectively for light and heavier pseudoscalar mesons.

As discussed above, in order to unveil the meson's properties at any larger scale, $\zeta > \zeta_H$, one needs to consider QCD all-orders evolution to account for the splitting of valence quasiparticles into more partons when the resolving scale increases\,\cite{Cui:2020tdf,Raya:2021zrz,Cui:2021mom,Cui:2022bxn,Lu:2022cjx,Yin:2023dbw}.

\begin{table}[h]
\caption{DA's moments $\langle x \rangle_{\varphi_\mathsf{P}}$ obtained at the hadron scale by implementing in Eq.\,\eqref{eq:DAfromBSA} the Bethe-Salpeter and gap equations' solutions with the quark + quark interaction kernel given by Eqs.\,(\ref{eq:kernelRL},\ref{eq:GQC}).}
\label{tab:DAmoments}
\centering
\begin{tabular}{c|ccc}
   $n \backslash \mathsf{P}$  &  $\eta_s$ & $\eta_c$ & $\eta_b$ \\ 
   \hline
    1 &  0.5002 & 0.5001 & 0.5000 \\
    2 &  0.3055 & 0.2847 & 0.2710 \\
    3 &  0.2082 & 0.1769 & 0.1566 \\
    4 &  0.1523 & 0.1165 & 0.0957 \\
    5 &  0.1172 & 0.0796 & 0.0617 \\
    6 &  0.0915 & 0.0589 & 0.0393 \\ 
    7 &  0.0699 & 0.0507 & 0.0208 \\
    \hline
\end{tabular}
\end{table}

\begin{table}[h]
\caption{The same as in Tab.\,\ref{tab:DAmoments}, herein for PDF's moments $\langle x^n\rangle_{q_\mathsf{P}}$ at the resolving hadron scale.}
\label{tab:PDFmoments}
\centering
\begin{tabular}{c|ccc}
   $n \backslash \mathsf{P}$  &  $\eta_s$ & $\eta_c$ & $\eta_b$ \\ 
   \hline
    1 &  0.5002 & 0.5003 & 0.5003 \\
    2 &  0.2846 & 0.2662 & 0.2572 \\
    3 &  0.1768 & 0.1492 & 0.1356 \\
    4 &  0.1172 & 0.0874 & 0.0732 \\
    5 &  0.0817 & 0.0532 & 0.0404 \\
    6 &  0.0593 & 0.0336 & 0.0228 \\ 
    7 &  0.0445 & 0.0191 & 0.0132 \\
    \hline
\end{tabular}
\end{table}

\begin{table}[ht]
    \centering
    \begin{tabular}{c|cc|cc|c}
         & $n_\varphi^\mathsf{P}$ & $\rho_\varphi^\mathsf{P}$ & $n_{\mathpzc q}^\mathsf{P}$ & $\rho_{\mathpzc q}^\mathsf{P}$ & $\alpha_\mathsf{P}$\\ \hline
         $\eta_c$ & 1.753 & 0.169 & 0.238 & 0.0468 & -20.6 \\ 
         $\eta_b$ & 0.0651 & 0.0484 & 0.0000516 & 0.01765 & -31.4 \\
         \hline 
    \end{tabular}
    \caption{Parameters for Eqs.\,\eqref{eq:ansatzsscc} as they result for the least-squares fit of their moments to those of Tabs.\,\ref{tab:DAmoments} and \ref{tab:PDFmoments}.}
    \label{tab:params}
\end{table}

Then, using the parameter set above described for the interaction, along with the procedure outlined in \cite{Xu:2025cyj} to extract the Mellin moments for both DAs according to \eqref{eq:DAfromBSA} and PDFs following \eqref{eq:DSEPDF}, one is left with the results collected in Tabs.\,\ref{tab:DAmoments} and \ref{tab:PDFmoments}. These moments are found to satisfy the physical constraints  \cite{Cui:2022bxn} shown in Eq.\,\eqref{eq:constraints}, as expected for DAs and PDFs of quarkonia bound-states calculated at the hadron scale. A hierarchic behavior of moments, their value increasing from light to heavy mesons, is apparent from Tabs. \ref{tab:DAmoments} and \ref{tab:PDFmoments}. Specialized for $n=2$, this translates into 
\begin{align}
\centering
\begin{tabular}{cccc}
     & \!\!\!\!$\eta_s$ & \!\!\!\!$\eta_c$ & $\eta_b$  \\
   $w_{\varphi_\mathsf{P}}$:  
   & 0.0555 $\;>$ & 0.0347 $\;>$& 0.0210 \\
   $w_{q_\mathsf{P}}$:  &  0.0346 $>$ & 0.0162 $>$ & 0.0072
\end{tabular}
\label{widths}
\end{align}
with $w=\langle (1-2x)^2/4 \rangle = \langle x^2 \rangle - 1/4$, directly exposing the width of the distribution around $x=1/2$. Note further that $w_{\varphi_\mathsf{P}} > w_{q_\mathsf{P}}$, as discussed below Eqs.\,\eqref{eq:ratios} in connection with the LFWF modification \eqref{eq:modif}. 

Following \emph{e.g.} Ref.\,\cite{Raya:2024ejx}, the moments reported in Tabs. \ref{tab:DAmoments} and \ref{tab:PDFmoments} can be used for the reconstruction of DAs and PDFs by modeling the distributions with the following ans\"atze: 
%
%
%
\begin{subequations}\label{eq:ansatzsscc}
\begin{align}
\varphi_\mathsf{P}(x) &= n_\varphi^\mathsf{P} x(1-x) 
\exp{\left(\frac{x(1-x)}{\rho_\varphi^\mathsf{P}}\right)} \;, \\
{\mathpzc q}_\mathsf{P}(x)  &= n_{\mathpzc q}^\mathsf{P} x^2(1-x)^2 
\exp{\left( \frac{x (1-x)}{\rho_{\mathpzc q}^\mathsf{P}} \right)}   \;;    
\end{align}
\end{subequations}
for $\mathsf{P}$=$\eta_c, \eta_b$. The parameters are obtained by a least-square fit of the moments obtained with \eqref{eq:ansatzsscc} to those from Tabs.\,\ref{tab:DAmoments} and \ref{tab:PDFmoments}, and can be found in Tab.\,\ref{tab:params}. The reconstructed curves are shown in the left plots of Fig.\,\ref{fig:DFvsDA2heavy}. 

In the case of $\eta_s$, fully in line with the conclusions presented in Ref.\,\cite{Yao:2025xjx}, the moments collected in Tabs.\,\ref{tab:DAmoments} and \ref{tab:PDFmoments} are compatible with asymptotic DA and PDF: 
$\varphi_{\eta_s}(x)=6x(1-x)$, ${\mathpzc q}_{\eta_s}=30 x^2(1-x)^2$; both agreeing with Eq.\,\eqref{eq:PDFeqDA2}, connecting the pseudoscalar meson's DA and PDF in the light quark sector.   

For the heavier pseudoscalar systems, the ratios of LFWF Mellin moments \eqref{eq:ratios} are displayed in the right panels of Fig.\,\ref{fig:DFvsDA2heavy} and, as highlighted above in Sec.\,\ref{sec:nonsep}, are shown to behave according to a qualitative pattern consistent with the effect of Eq.\,\eqref{eq:modif} made apparent in Eqs.\,\eqref{eq:momentstilda} and \eqref{eq:intR0}.  
Then, in order to apply Eq.\,\eqref{eq:chi2} and to follow the quantitative analysis described in connection with it, one needs to define first a parametric functional form $\kappa_{\alpha_\mathsf{P}}(x)$ respecting the symmetry $x \leftrightarrow 1-x$ and obeying\footnote{Preserving the endpoints' behavior requires $\kappa_{\alpha_\mathsf{P}}(x) \to c$ as $x\to 0,1$; where $c$ is a constant, the same at both endpoints for symmetric reasons, which can be simply borrowed by PDF normalization.} $\kappa_{\alpha_\mathsf{P}}(x) \to 1$ when $x \to 0,1$. A good candidate is
\begin{align}\label{eq:kappa}
\kappa_{\alpha_\mathsf{P}}(x) = \exp{\left(\alpha_\mathsf{P} \, x^2(1-x)^2\right)} \;,    
\end{align}
which, among a few other ans\"atze\,\footnote{We have tried different functions of the argument $x(1-x)$,  as $1+\alpha x^m (1-x)^m$ or $\exp{\left(\alpha x^m (1-x)^m \right)}$ for different values of $m$.}, has proven to deliver the best least-squares fits of moments. The best-fit estimates for $\alpha_\mathsf{P}$ for $\mathsf{P}=\eta_c,\eta_b$ can be found in Tab.\,\ref{tab:params}. Therefore, replacing $\kappa(x)$ with $\kappa_{\alpha_\mathsf{P}}(x)$ in Eq.\,\eqref{eq:q5vsDA2}, evaluated for the best-fit values of $\alpha$, the PDFs are obtained and displayed with a red dashed line in the right panels of Fig.\,\ref{fig:DFvsDA2heavy}, derived only from the Mellin moments of LFWFs and the prescription \eqref{eq:modif}. As can be seen, they show an excellent agreement with the alternative outcomes from Eq.\,\eqref{eq:DSEPDF}, drawn with a blue solid line, in both $\eta_c$ and $\eta_b$ cases. This agreement can be understood as strongly supporting the validity of the prescription \eqref{eq:modif}, confirmed by the use of CSM as a remarkable approximation.

All in all, the reconstruction of the CSM Mellin moments gathered in Tabs.\,\ref{tab:DAmoments} and \ref{tab:PDFmoments} produces the expected distribution profiles, shown in the right panels of Fig.\,\ref{fig:DFvsDA2heavy}, peaking around $x=1/2$ and the width shrinking with increasing meson mass. The latter is recalling that the probability for a quasiparticle carrying a given lightcone momentum fraction becomes the more and more localized within the valence domain, the larger is its mass. This is also quantitatively exposed in \eqref{widths}, which shows that the DA/PDF width is reduced by approximately a factor 1.7/2.3, when replacing charm by bottom quarks. The stronger narrowing for the PDF is consistent with Eq.\,\eqref{eq:q5vsDA2}, and can be explained by the combination of both the squaring of the DA and the further PDF compression induced by $\kappa(x) \to \kappa_{\alpha_\mathsf{P}}(x)$, which also increases with the bound-state mass.

\section{An illustrative LFWF model}
\label{sec:illustration}

\begin{figure}[htbp]
    \centering
    \begin{tabular}{cc}
        \includegraphics[width=0.45\textwidth]{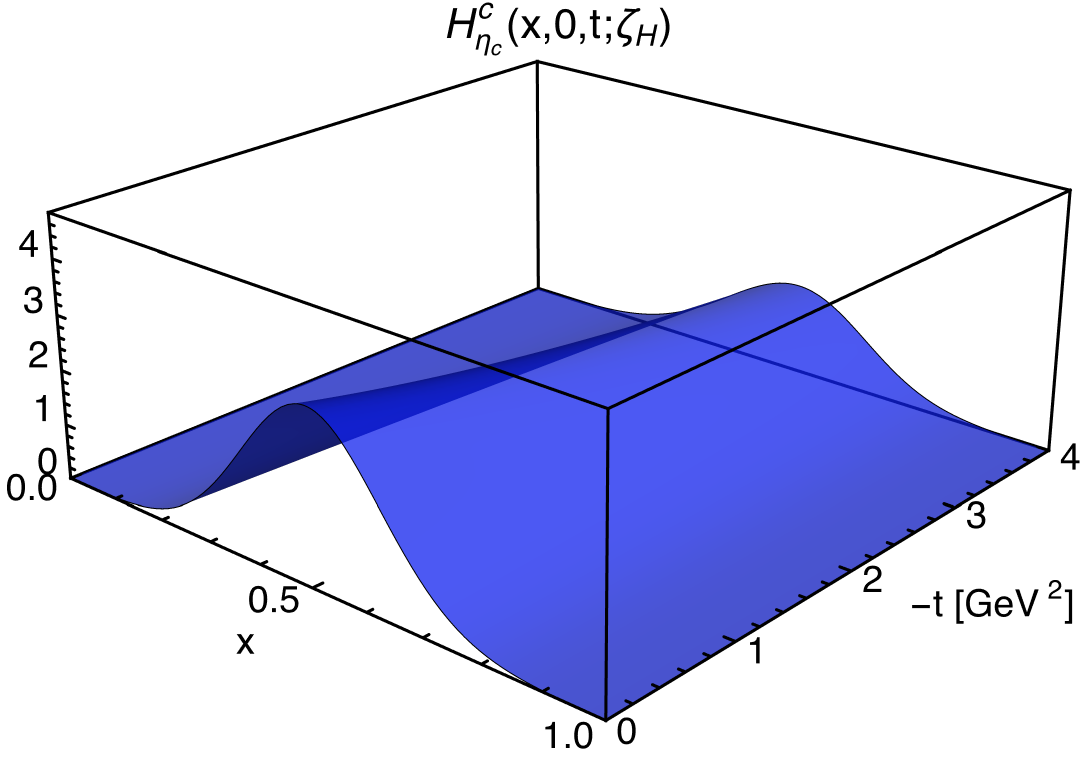}
        &
        \hspace{0.05\textwidth}
        \includegraphics[width=0.45\textwidth]{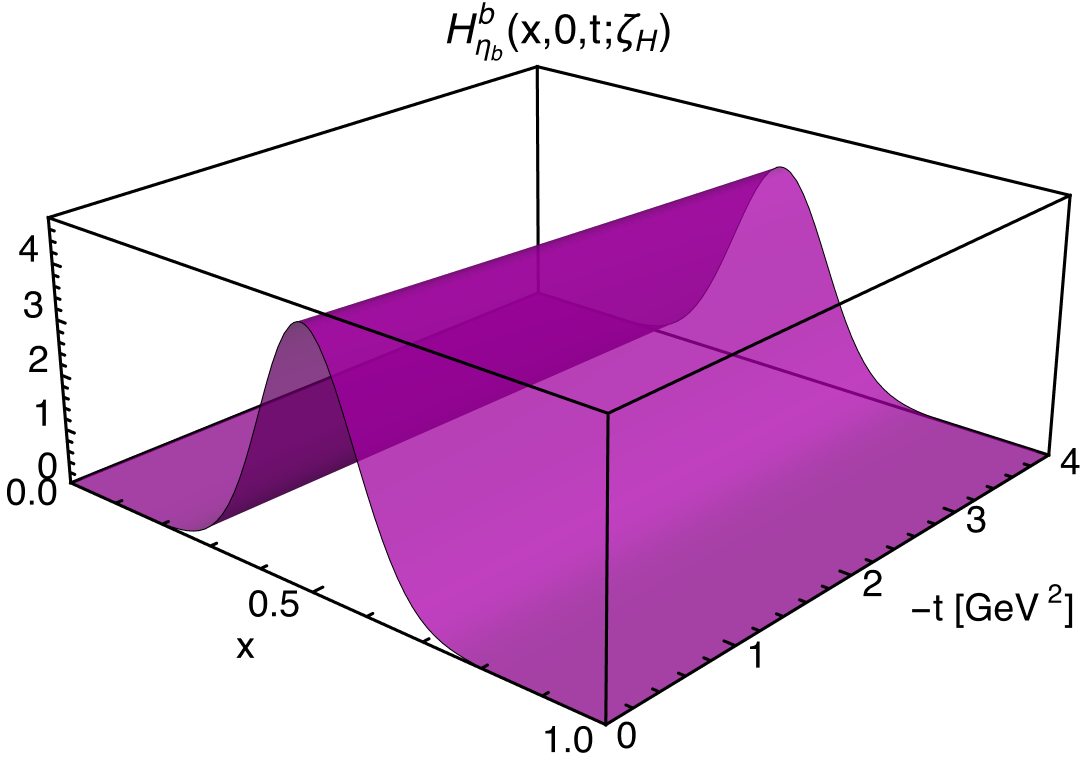}
    \end{tabular}
    \caption{Zero-skewness GPDs, calculated from Eq.~\eqref{eq:tGPD}
for $\textbf{P}=\eta_c$ (left panel) and $\textbf{P}=\eta_b$ (right), at the hadron scale $\zeta_H$.}
    \label{fig:tGPDs}
\end{figure}


   \begin{figure}[htbp]
    \centering
    \begin{tabular}{cc}
        \includegraphics[width=0.45\textwidth]{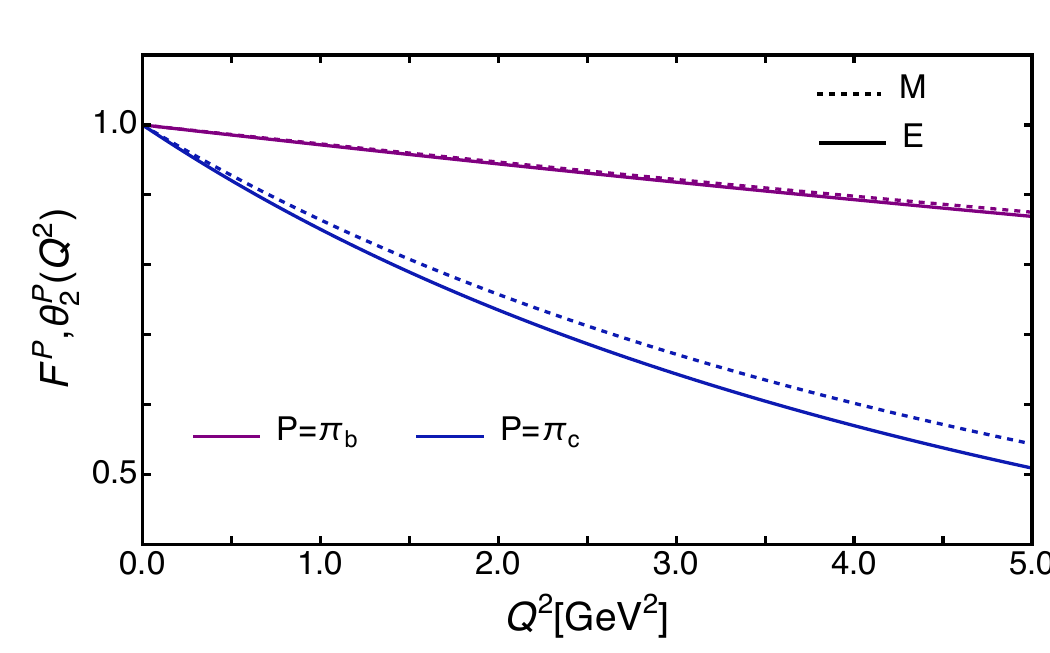}
        &
        \hspace{0.05\textwidth}
        \includegraphics[width=0.45\textwidth]{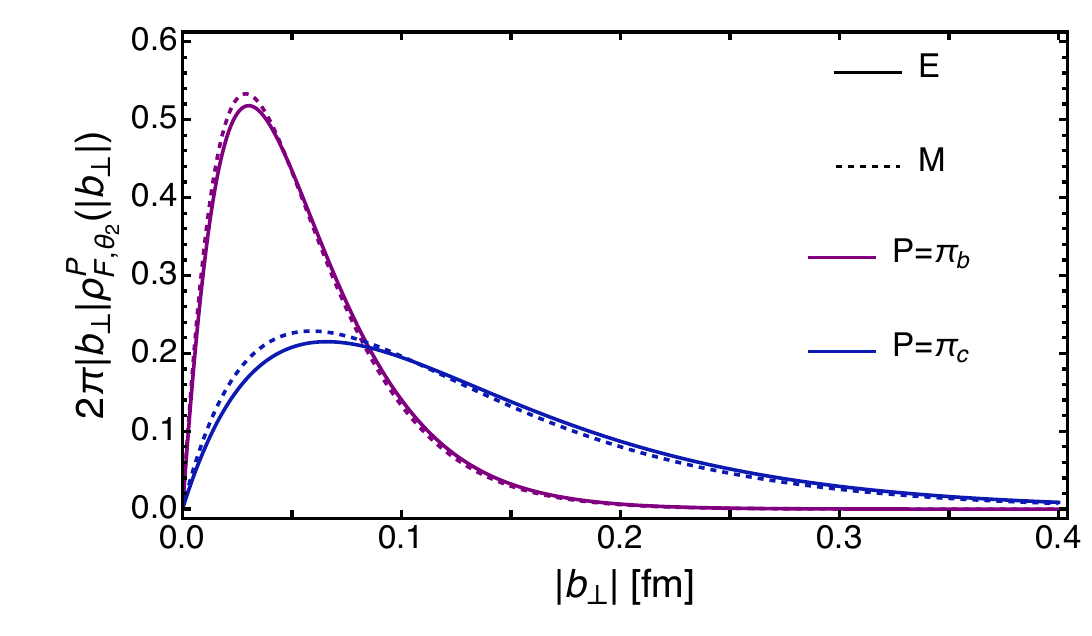}
    \end{tabular}
    \caption{Left panel: Electromagnetic and mass form factors.
Solid curves -- the electromagnetic form factors of the fictitious charged systems
$\pi_c$ and $\pi_b$; dashed curves -- the corresponding gravitational form factors.
Right panel: Transverse charge and mass density distributions.
Purple and blue curves correspond to $\pi_b$ and $\pi_c$, respectively.
For the mass form factors and mass densities, the results are identical for
$\eta_{c,b}$ and the corresponding fictitious charged systems $\pi_{c,b}$.}
    \label{fig:FFs}
\end{figure}

The prescription\,\eqref{eq:modif}, which extends here the factorised form for light pseudoscalar meson LFWFs\,\cite{Yao:2025xjx} to the case of charmonium and bottomonium, is also considered\,\footnote{In Ref.\,\cite{Raya:2026gwg}, the proposed modification is $M_q \to {\cal M}(x)$, redefining there our compression function here as $\kappa(x)={\cal M}^2(x)/M_q^2$.} in Ref.\,\cite{Raya:2026gwg} and applied there to a given PTIR algebraic model for the pion. This model corresponds to a generalization of a previous, insightful one introduced in Ref.\,\cite[App.\,A]{Raya:2021zrz} and further applied to successfully describe the DVCS beam spin asymmetry\,\cite{Chavez:2021koz,Chavez:2021llq}. In this latter particular model, the pseudoscalar pion LFWF is fixed by the following choices: $F^0(k_\perp^2)=M_q^2 / [k_\perp^2+M_q^2]^2$ and $F^1(k_\perp^2)\equiv 0$. Thus, applying Eq.\,\eqref{eq:modif} with $\kappa(x) \to \kappa_{\alpha_\mathsf{P}}(x)$, one is left with  
\begin{align}
\label{eq:factLFWFf}
\psi_\mathsf{P}^0(x,k_\perp^2) = 16 \pi^2 f_\mathsf{P} N^{1/2}_{\mathpzc q}
\frac{M_q^2 \kappa(x)}{\left[k_\perp^2+M_q^2 \kappa_{\alpha_\mathsf{P}}(x)\right]^2} \, 
\varphi_\mathsf{P}(x) \;,    
\end{align}
for $\mathsf{P}=\eta_c,\eta_b$; and, according to Eq.\,\eqref{DefPDF}, 
\begin{align}\label{eq:qPDF}
{\mathpzc q}_\mathsf{P}(x) = N_{\mathpzc q} \frac{\varphi^2_\mathsf{P}(x)}{\kappa_{\alpha_\mathsf{P}}(x)} \;,    
\end{align}
as expected, where $N_{\mathpzc q}$ is introduced to give the correct PDF normalization guaranteeing baryon number conservation.
In this section, this non-separable, extended version of the LFWF model from Ref.\,\cite[App.\,A]{Raya:2021zrz} will be used to illustrate how one can capitalize on the CSM results shown in the previous section to sketch some internal structure properties for pseudoscalar quarkonia. It should be implicitly understood that the LFWF in Eq.\,\eqref{eq:factLFWFf} and all the results derived from it, as Eq.\,\eqref{eq:qPDF} and those which follow below, are obtained at the hadron scale, $\zeta_H$.  

Indeed, as outlined in the appendix \ref{app:GPD}, one can rely on the so-called overlap representation\,\cite{Burkardt:2002hr,Diehl:2003ny} to prove that the zero-skewness quasiparticle meson-$\mathsf{P}$ GPD reads\footnote{
GPDs are extensively and instructively discussed, \emph{e.g.}, in Refs.\,\cite{Belitsky:2005qn,Mezrag:2022pqk,Mezrag:2023nkp}.}
\begin{align}\label{eq:tGPD}
H_\textbf{P}^q(x,0,t) = \, 
\theta(x) \, {\mathpzc q}_\textbf{P}(x) \, \Phi_\textbf{P}\left(\frac{s}{M_q^2}\right) \;,  
\end{align}
when the LFWF is given by Eq.\,\eqref{eq:factLFWFf}; where $t=-\Delta^2$ denotes the momentum transfer, with 
\begin{align}
\Phi_\textbf{P}(z) &= \frac{6}{\displaystyle \left(z + 4\right)^3} 
\left( 10 + z + \frac{8(z+1)}{z} \left[ 
\sqrt{\frac{z+4}{z}} \atanh{\left(\sqrt{\frac{z}{z+4}}\right)} -1\right]
\right) \;;      
\end{align}
and
\begin{align}\label{eq:s}
s=\frac{\Delta^2(1-x)^2}{\kappa_{\alpha_\mathsf{P}}(x)}    
\end{align}
fully encoding the correlation of $k_\perp$ and $x$ in the LFWF. Therefore, simply attaching the LFWF to the DA as given in Eq.\,\eqref{eq:factLFWFf}, a closed, analytic expression for the zero-skewness GPD can be delivered, and used to derive the charge and mass distributions within the meson from the corresponding form factors.      

To this last goal, leveraging on the polynomiality property of GPDs, one can directly calculate the valence-quark or quasiparticle contribution to the electromagnetic form factor from the zero-skewness GPD, thus reading
\begin{align}\label{eq:FFf}
F^q_\mathsf{P}(\Delta^2) = \int_{-1}^1 dx H_\textbf{P}^q(x,0,-\Delta^2)  = \int_0^1 dx \mathpzc{q}_\mathsf{P}(x) \Phi_\textbf{P}\left(\frac{\Delta^2(1-x)^2}{M_q^2 \kappa_{\alpha_\mathsf{P}}(x)} \right) \;;  
\end{align}
and, equivalently, for the gravitational form factor 
\begin{align}
\theta_2^{q_\textbf{P}}(\Delta^2) = \int_{-1}^{1} dx \, x H_\textbf{P}^q(x,0,-\Delta^2) = \int_0^1 dx \, x \, \mathpzc{q}^\textbf{P}(x) \Phi_\textbf{P}\left(\frac{\Delta^2(1-x)^2}{M_q^2 \kappa_{\alpha_\mathsf{P}}(x)} \right) \;,
\label{eq:theta2}
\end{align}
particularly attached to the mass distribution inside the meson\,\cite{Polyakov:2018zvc}; both relying on the LFWF given by Eq.\,\eqref{eq:tGPD}.

While the meson gravitational form factor reads $\theta_2^{\mathsf{P}}(\Delta^2)=2\theta_2^{q_\mathsf{P}}(\Delta^2)$, a definition for a meson electromagnetic form factor from the quasiparticle contributions is meaningless for $\eta_c$ or $\eta_b$, as they correspond to neutral bound states in which the two quasiparticle contributions exactly cancel each other. One can consider instead the fictitious $\pi^+_c$ and $\pi^+_b$ bound-state systems, made respectively by two valence-quarks from a first flavor generation with degenerate charm or bottom masses. It should be noted that, in our approach, the quasiparticle contributions to the form factors are the same for $\pi^+_{c,b}$ and $\eta_{c,b}$. They just combine differently, weighted by different quark electric charges, to produce the full meson electromagnetic form factors. In the case of $\pi^+_{c,b}$, as in pion's case, $F_\mathsf{\pi_{c,b}}(\Delta^2) = F_\mathsf{\pi_{c,b}}^{c,b}(\Delta^2)$. 

Thus, we can define the mass and charge density-distributions in the light-front transverse plane ($\vec{b}_\perp$ denotes the 2-dimensional vector in the plane) as follows\,\cite{Miller:2018ybm,Burkardt:2002hr,Miller:2010tz}
\begin{align}\label{eq:rhoP}
\rho^\mathsf{P}_{\{F,\theta_2\}}(|b_\perp|) = \int_0^\infty \frac{d\Delta}{2\pi} \Delta J_0(\Delta |b_\perp|) \left\{ F_\mathsf{P}(\Delta^2),\theta_2(\Delta^2)\right\} \;,   
\end{align}
while the radii for these distributions, $\rho_\mathsf{P}^{e,m}$, and for the corresponding 3-dimensional ones, $r_\mathsf{P}^{e,m}$, read\footnote{
Although beyond the scope of the present work, it is worth mentioning that the choice between either 3-dimensional\,\cite{Hofstadter:1958psf,Ernst:1960zza,Sachs:1962zzc} or 2-dimensional transverse distributions to produce a spatial image of a hadron has recently become controversial\,\cite{Jaffe:2020ebz,Freese:2021mzg,Epelbaum:2022fjc,Chen:2022smg,Lorce:2022jyi,Lorce:2020onh,Chen:2023dxp}
}
\begin{align}\label{eq:r2p} 
\left( 
\begin{array}{c}
\{\left(\rho^e_\mathsf{P}\right)^2,\left(r^e_\mathsf{P}\right)^2\} \\ 
\{\left(\rho^m_\mathsf{P}\right)^2,\left(r^m_\mathsf{P}\right)^2\} 
\end{array}
\right)= - \{4,6\} \left. \frac{d}{d(\Delta^2)} 
\left( \begin{array}{c}
F_\mathsf{P}(\Delta^2) \\ 
\theta_2^\textsf{P}(\Delta^2) 
\end{array}
\right) \right|_{\Delta^2=0}
= \frac {\{12,18\}}{5 M_q^2} \left( 
\begin{array}{c}
\langle x^2/\kappa_{\alpha_\mathsf{P}}(x)  \rangle_{{\mathpzc q}_\mathsf{P}} \\ 
\langle 2 x^2 (1-x)/\kappa_{\alpha_\mathsf{P}}(x)  \rangle_{{\mathpzc q}_\mathsf{P}}
\end{array}
\right) \;. 
\end{align}
In both 2- and 3-dimensional cases, one can derive the following result for the ratios of mass to charge radii 
\begin{align}\label{eq:rmvsrq}
\frac{\left(\rho_\mathsf{P}^m\right)^2}{\left(\rho_\textbf{P}^e\right)^2} =  \frac{\left(r_\mathsf{P}^m\right)^2}{\left(r_\textbf{P}^e\right)^2} =
\frac{2 \langle x^2(1-x) \rangle_{\widetilde{q}_\mathsf{P}}}{\langle x^2 \rangle_{\widetilde{q}_\mathsf{P}}} \;.
\end{align}
In the case of the mass radii, $\rho^m_\mathsf{P}$ and $r^m_\mathsf{P}$  denote indistinctly either $P=\eta_{c,b}$ or $P=\pi^+_{c,b}$, taking  the same value in both cases; by contrast, $\rho^e_\mathsf{P}$ and $r^e_\mathsf{P}$ are only meaningful for $P=\pi^+_{c,b}$, and consequently this also applies to the ratios \eqref{eq:rmvsrq}. The averages $\langle \, \cdot \,\rangle_{{\mathpzc q}_\mathsf{P}}$ and $\langle \, \cdot \,\rangle_{\widetilde{q}_\mathsf{P}}$ are evaluated for the distributions ${\mathpzc q}_\mathsf{P}(x)$ and $\widetilde{q}_\mathsf{P}(x)=\widetilde{N}_\mathsf{P} {\mathpzc q}_\mathsf{P}(x)/\kappa_{\alpha_\mathsf{P}}(k)$, respectively, both properly normalized to unity when $\widetilde{N}_\mathsf{P}$ is correctly defined. Eq.\,\eqref{eq:rmvsrq} generalizes the result given in Ref.\,\cite[Eq.\,(41)]{Raya:2021zrz} for factorized LFWFs, which can be simply recovered by replacing $\kappa_{\alpha_\mathsf{P}}(x) \to 1$. As it is therein demonstrated, as far as the distribution $\widetilde{q}_\mathsf{P}$ is properly normalized, Eq.\,\eqref{eq:rmvsrq}'s rhs is upper bounded by unity,  and hence, $r^m_\mathsf{P} \leq r^e_\mathsf{P}$, the narrower around $x$=1/2 is the distribution the closer to unity [it gives exactly 1 when $\widetilde{q}_\mathsf{P}(x) = \delta(x-1/2)$]. 

As shown in Tab.\,\ref{tab:params}, the value of $-\alpha_\mathsf{P}$ increases with the bound-state mass, thereby strengthening the compressing effect of $k_{\alpha_\mathsf{P}}(x)$ and driving the ratio of mass to charge radii given by\,\eqref{eq:rmvsrq} toward unity, \emph{i.e.}, the non-relativistic limit of an infinitely heavy meson. This trend is clearly exposed in the fourth row of Tab.\,\ref{tab:radii}.    
 
The results of Eq.\,\eqref{eq:rmvsrq} for the ratios of radii depend solely on the CSM outputs presented in Sec.\,\ref{sec:DSEresults} and on the corresponding determination of the compressing function $\kappa_{\alpha_\mathsf{P}}(x)$. However, obtaining the radii themselves, as well as the form factors and density distributions in Eqs.\,(\ref{eq:FFf}-\ref{eq:rhoP}), requires fixing $M_q$ (see Eq.\,\eqref{eq:factLFWFf}) as the only additional parameter. To this purpose, we replace $M_q$ with the following dressed quark masses: $M_c=1.50$ GeV and $M_b=4.50$ GeV, in fair agreement with, \emph{e.g.}, Ref.\,\cite[Tab.II]{Almeida-Zamora:2023bqb}, where algebraic LFWF models similar to Eq.\,\eqref{eq:factLFWFf} have been used to describe heavy-light pseudoscalar mesons. 

Interestingly, as it is shown in Tab.\,\ref{tab:radii}, the electromagnetic radii for $\pi_c$ and $\pi_b$ can be obtained through Eq.\,\eqref{eq:r2p} and seen to agree with the CSM results presented in Ref.\cite[Tab.II]{Xu:2024vkn}. More importantly, this agreement strongly relies on the effect of the compressing function $k_{\alpha_\mathsf{P}}(x)$ in Eq.\,\eqref{eq:r2p}, as can be seen by replacing it with unity, thereby recovering the separable LFWF model. In that case, radii are reduced by approximately a factor two, strongly disagreeing by a 50 \% with the estimates from Ref.\,\cite{Xu:2024vkn}, which are obtained from electric charge distributions within a CSM framework. Thus, the prescription \eqref{eq:modif}, which amends the separable LFWF model and, eventually, introduces the compressing factor $\kappa_{\alpha_\mathsf{P}}(x)$ in Eq.\,\eqref{eq:r2p}, is also well supported by this crosscheck of the radii.         

\begin{table}
\begin{tabular}{r|ccc}
$\mathsf{P}$ \; & $\pi_s$ & $\pi_c$ & $\pi_b$ \\
\hline
$\kappa_{\alpha_\mathsf{P}}\equiv$ \eqref{eq:kappa} \; & 0.443 & 0.227 & 0.107 \\ 
$\kappa_{\alpha_\mathsf{P}}\equiv 1$ \; & 0.443 & 0.129 & 0.042 \\
Ref.\,\cite{Xu:2024vkn} \; & 0.466 & 0.227 & 0.113 \\
\hline
Eq.\,\eqref{eq:rmvsrq} \; & 0.750 & 0.904 & 0.954 
\end{tabular}
\caption{The three first rows deliver the three-dimensional electromagnetic radii (in fm) calculated according to Eq.\,\eqref{eq:r2p} with the compressing function $\kappa_{\alpha_\mathsf{P}}(x)$ given by Eq.\,\eqref{eq:kappa} and the parameter $\alpha_\mathsf{P}$ from Tab.\,\ref{tab:params} (first row), with $\alpha_\mathsf{P}=0$ such that a separable LFWF model is recovered (second) and, for the sake of comparison, the outputs presented in Ref.\,\cite[Tab.II]{Xu:2024vkn} (third). The two-dimensional outputs, as it is apparent in Eq.\,\eqref{eq:r2p}, can be simply obtained by applying a factor 2/3 to all these entries. The fourth row shows the ratios of mass to charge radii, as given by Eq.\,\eqref{eq:rmvsrq}, from which mass radii can be immediately obtained. The case of a bound-state of two first-generation valence-quarks with degenerate strange mass, $\pi_s$, is herein included as a benchmark from separable LFWF models, and evaluated with the asymptotic DA, ${\mathpzc q}_{\pi_s}=30 x^2(1-x)^2$, $\alpha_{\pi_s}=0$ and $M_s=0.450$ GeV.
}
\label{tab:radii}
\end{table}

Then, crossreferencing Eqs.\,(\ref{eq:tGPD}-\ref{eq:s}), we can calculate the zero-skewness GPDs depicted in Fig.\,\ref{fig:tGPDs}; and the form factors (left panels) and density-distributions (right) displayed in Fig.\,\ref{fig:FFs} from Eqs.\,(\ref{eq:FFf}-\ref{eq:rhoP}). 

The hadron-scale GPDs in Fig.\,\ref{fig:tGPDs} show the same qualitative picture sketched in Ref.\,\cite{Mezrag:2014jka} for the pion and confirmed by a plethora of further works, \emph{e.g.}\, Refs.\,\cite{Burkardt:2015qoa,Mezrag:2016hnp,Chouika:2017dhe,Chouika:2017rzs,Xu:2018eii,Zhang:2021mtn,Raya:2021zrz,Chavez:2021llq,Chavez:2021koz,Albino:2022gzs,Xing:2023eed}; namely, that the maximum is located at $\Delta^2=0$ (where GPD and PDF coincide) and $x=1/2$ (as far as the quarks masses remain degenerate); and that the position of this peak shifts toward $x=1$, its height diminishes and the profile around the peak narrows as $\Delta^2$ increases, recalling that the harder is the probe used to reveal the meson structure the more it focuses on the valence domain. The integrated zero-skewness GPD and the position of its peak at each fixed value of $\Delta^2$, see Eqs.\,(\ref{eq:FFf},\ref{eq:theta2}) and the left panel of Fig.\,\ref{fig:FFs}, relate to the electromagnetic and gravitational form factors, respectively, which decrease with increasing momentum transfer.      

From light to heavy mesons, the profile also becomes narrower at $\Delta^2=0$, as the PDF, and the rate at which both the peak height and the profile width decrease with increasing $\Delta^2$ becomes slower. This is transparent from Eqs.\,(\ref{eq:FFf}-\ref{eq:s}), which make the two following kinematical limits very apparent: 
\begin{itemize}
\item (i) as $\Delta^2 \gg M_q^2$, $s/M_q^2 \to \infty$ and hence $\Phi_\mathsf{P} \to 0$ in \eqref{eq:FFf}, except when $x \to 1$ such that $s/M_q^2 \to 0$ and $\Phi_\mathsf{P} \to 1$;  
\item and (ii) $\Delta^2 \ll M_q^2$, implying that $s/M_q^2 \to 0$ and $\Phi_\mathsf{P} \to 1$ in \eqref{eq:FFf} $\forall x$. 
\end{itemize}
The first limit explains why the GPD peak shifts toward $x=1$ and the width narrows when $\Delta^2$ increases; the second why the rate at which height and width shrink becomes slower as the mass increases, since $H_\mathsf{P}^q(x,0,-\Delta^2) \simeq {\mathpzc q}_\mathsf{P}(x)$ for $\Delta^2 \ll M_q^2$. In both cases, the role played by $\kappa_{\alpha_\mathsf{P}}(x)$ is that of increasing $s$ from Eq.\,\eqref{eq:s}, especially around $x=1/2$, interfering positively in case (i) and negatively in (ii); \emph{i.e.}, bolstering the effect of the limit (i) and smoothing that of (ii). This impact of $\kappa_{\alpha_\mathsf{P}}(x)$ is also consistent with that shown in Tab.\,\ref{tab:radii}: comparing the first and second rows, it becomes clear that the presence of $\kappa_{\alpha_\mathsf{P}}(x)$ in Eq.\,\eqref{eq:r2p} is roughly doubling the zero-momentum negative slope of the form factors given by Eqs.\,(\ref{eq:FFf},\ref{eq:theta2}), \emph{i.e.}, roughly doubling their rate of decrease. 

This rate of decrease for the electromagnetic and gravitational form factors, depicted in the left panel of Fig.\,\ref{fig:FFs}, is seen to be softened by the increasing meson mass. The heavier a meson the harder its form factors profiles, entailing that the corresponding distributions of mass and charge are more localized. The latter is illustrated by the density distributions (see the right panel of Fig.\,\ref{fig:FFs}), where the profile density of the $\eta_b$ lies deeper inside, closer to the center of transverse momentum, than that of $\eta_c$. For each meson, when comparing mass and charge form factors and distribution-densities, these tend to converge as one approaches the non-relativistic limit of an infinitely heavy meson, as shown previously in the last row of Tab.\,\ref{tab:radii}, as a result of Eq.\,\eqref{eq:rmvsrq}. As can be seen in both panels of Fig.\,\ref{fig:FFs}, in the case of the $\eta_b$, both the form factors and the density distributions for mass and charge almost coincide.

%

\section{Conclusions}
\label{sec:conclusions}

In this note, we explored the internal structure of heavy pseudoscalar mesons, specifically charmonium ($\eta_c$) and bottomonium ($\eta_b$), by computing and analysing their DAs and PDFs, and in particular how they are connected by their underlying LFWFs. We build upon previous studies of lighter mesons, such as the pion and kaon, to determine how increasing meson mass affects the mathematical structure of the LFWFs and the resulting distributions.

We employed CSM, \emph{i.e.}, we solved the corresponding DSEs and BSEs to calculate the BSWFs and, from them, we extracted the $k_\perp$-dependent Mellin moments of the LFWFs. The scrutiny of of these Mellin moments reveals the breakdown of the separable (factorized) form of the LFWF that typically works for light mesons. In the latter case, the LFWF dependence on lightcone and transverse momentum decouple, entailing that the ratios of $m$-th to $0$-th order moments behave as a constant in terms of transverse momenta. In the case of heavier mesons, a well-defined deviation from this constant profile for these ratios indicates the coupling of the dependences on lightcone and transverse momenta. This can be simply accounted by introducing a given function of the lightcone momentum, $\kappa(x)$, that modifies the mass dimension in the LFWF. This minimal modification allows for a suitable  extension of light-meson separable models to heavier systems, describing properly the ratios of CSM LFWF moments, while maintaining physical consistency.

The DAs and PDFs obtained from the reconstruction of their Mellin moments, calculated within the CSM framework, are significantly narrower for $\eta_b$ than for $\eta_c$ or light mesons as the pion and kaon or fictitious bound-states of light quarks as $\eta_s$ or $\pi_s$. This tells that the probability for a quasiparticle to carry a specific momentum fraction becomes more localized, as the mass of the meson increases. This mass-induced compression is already apparent for the DAs, and appears enhanced at the level of the PDFs. Such enhancement can be explained as a direct consequence of the introduction of $\kappa(x)$ when extending any light-meson separable model to heavier systems. Relying only on this minimal modification, the ratio of mass to charge radii can be calculated, also extending to heavier systems a previous expression for light mesons. This ratio is found to be smaller than, but tending to, unity when the meson mass increases and the non-relativistic limit of an infinitely heavy meson is approached. 

Finally, after estimating this compression function $\kappa(x)$ by comparing the CSM outputs for DAs and PDFs, we applied it to extend a particular LFWF separable model, formerly employed to calculate the DVCS beam spin asymmetry, in order to illustrate its implications on the meson internal structure. To this goal, we calculated the zero-skewness quasiparticle GPDs for the $\eta_c$ and $\eta_b$, and, from them, calculated their electromagnetic and gravitational form factors and the corresponding transverse-plane density distributions. The resulting picture is clear and consistent with previous findings. Namely, both the electromagnetic and gravitational form factors show a decrease with momentum transfer that weakens with increasing meson mass, the profile of the latter always being harder than that of the former, tending however to converge. Therefore, charge and mass densities lie deeper inside and more localized as the meson mass increases, as shown by the corresponding density profiles. The charge distributions extend farther than mass distributions but tend to coincide in the non-relativistic limit of an infinitely heavy meson.        

\medskip
\noindent\textbf{Acknowledgments}.
Work supported by the Spanish Ministry of Science and Innovation (MICINN grant no.\ PID2022-140440NB-C22), and Junta de Andalucía (grant no.\ P18-FR-5057). Z.Q. Yao acknowledges support from Helmholtz-Zentrum Dresden-Rossendorf, under the High Potential Programme. J. R-Q is indebted to the program Jean d'Alembert which partially supported this work through a "Chair d'Excellence".

\medskip
\noindent\textbf{Data Availability Statement}. This manuscript has no associated data or the data will not be deposited. All information necessary to reproduce the results described herein is contained in the material presented above.

\appendix

\section{zero-skewness GPD}
\label{app:GPD}

Following Ref.\,\cite{Raya:2021zrz}, an approximated three-dimensional image of the meson $\mathsf{P}$ can be obtained by using the GPD overlap representation\,\cite{Burkardt:2002hr,Diehl:2003ny} only with helicity-0 components of LFWFs, neglecting helicity-1. The GPD (in the valence-quark or quasiparticle case at the hadron scale) reads
%
\begin{align}\label{eq:overlap}
H^q_\mathsf{P}(x,\xi,t) = \theta(x_-)  \int \frac{d^2{k_\perp}}{16 \pi^3}
\psi_{\mathsf{P}}^{0\ast}\left(x_-,{k}_{\perp -}^2 \right)
\psi_{\mathsf{P}}^{0}\left( x_+,{k}_{\perp +}^2 \right) \,,
\end{align}
%
with: $P=(p^\prime+p)/2$, where $p^\prime$, $p$ are the final, initial meson momenta in the defining scattering process;
$\Delta = p^\prime - p$, $P\cdot \Delta = 0$, $t=-\Delta^2$;
the ``skewness''
\begin{equation}
\xi = [-n\cdot \Delta]/[2 n\cdot P]\,,\;\mbox{$|\xi|\leq 1$}\,;
\end{equation}
and
\begin{subequations}
\begin{align}
x_\mp & = (x \mp \xi)/(1\mp \xi)\,, \\
{k}_{\perp \mp} & = {k}_\perp \pm ({\Delta}_\perp/2)(1-x)/(1 \mp \xi) \,.
\end{align}
\end{subequations}
It can be clearly seen that Eq.\,\eqref{DefPDF} follows from considering $\Delta=0$ (forward limit) in Eq.\,\eqref{eq:overlap} and neglecting the contribution of helicity-1 components. 

For our purpose of deriving form factors and charge and mass densities, without loss of information, we can focus in the following on the zero-skewness case (also implying that the momentum transfer becomes transverse: $\Delta^2=Q_\perp^2$). Then, specializing Eq.\,\eqref{eq:overlap} for $\xi=0$ and using the LFWF defined by Eq.\,\eqref{eq:factLFWFf}, one is left with the following expression for the zero-skewness GPD ($u$ is a standard Feynman parameter)
\begin{align}
H_\mathsf{P}^q(x,0,t) &= \theta(x) N_{\mathpzc q} \,\varphi_{\mathsf{P}}^2(x) \kappa_{\alpha_\mathsf{P}}^2(x) \,   \int_0^1 du \frac{6 u(1-u)}{\left( \kappa_{\alpha_\mathsf{P}}(x) + \displaystyle \frac{\Delta^2}{M_q^2} (1-x)^2 u(1-u)\right)^3}  \nonumber \\
&=  
\theta(x) \, {\mathpzc q}_\textbf{P}(x) \, \Phi_\textbf{P}\left(\frac{s}{M_q^2}\right) \;,
\label{eq:AptGPD}
\end{align}
with the DA is replaced by the PDF according to Eq.\,\eqref{eq:qPDF}, 
\begin{align}
s = \frac{\Delta^2 (1-x)^2}{\kappa_{\alpha_\mathsf{P}}(x)} \;    
\end{align}
and
\begin{align}
\Phi_\textbf{P}(z) &= \int_0^1 du \frac{6 u(1-u)}{\left(1+ \displaystyle z u(1-u)\right)^3} \nonumber \\ 
& =
\frac{6}{\displaystyle \left(z + 4\right)^3} 
\left( 10 + z + \frac{8(z+1)}{z} \right.  \left. \left[ 
\sqrt{\frac{z+4}{z}} \atanh{\left(\sqrt{\frac{z}{z+4}}\right)} -1\right]
\right) \;.   
\end{align}
Expanding $\Phi_\textbf{P}(z)$ around $z=0$, it reads
\begin{align}
\Phi_\textbf{P}(z) = 1 - \frac 3 5 z + {\cal O}(z^2) \;,     
\end{align}
what exposes the low-$\Delta^2$ behaviour of the meson-$\textsf{P}$ zero-skewness GPD and makes apparent that, in the forward limit, the PDF is recovered.

\bibliographystyle{elsarticle-num-names}
\bibliography{CollectedBiB}

@article{Miramontes:2026san,
    author = "Miramontes, A. S. and Morgado, J. M. and Papavassiliou, J.",
    title = "{Baryonic form factors of light pseudoscalar mesons}",
    eprint = "2604.00959",
    archivePrefix = "arXiv",
    primaryClass = "hep-ph",
    doi = "10.1016/j.physletb.2026.140531",
    journal = "Phys. Lett. B",
    volume = "878",
    pages = "140531",
    year = "2026"
}

@article{Broniowski:2024oyk,
    author = "Broniowski, Wojciech and Ruiz Arriola, Enrique",
    title = "{Gravitational form factors of the pion and meson dominance}",
    eprint = "2405.07815",
    archivePrefix = "arXiv",
    primaryClass = "hep-ph",
    doi = "10.1016/j.physletb.2024.139138",
    journal = "Phys. Lett. B",
    volume = "859",
    pages = "139138",
    year = "2024"
}

@article{Ydrefors:2021dwa,
    author = "Ydrefors, Emanuel and de Paula, Wayne and Nogueira, Jorge H. Alvarenga and Frederico, Tobias and Salm{\'e}, Giovanni",
    title = "{Pion electromagnetic form factor with Minkowskian dynamics}",
    eprint = "2106.10018",
    archivePrefix = "arXiv",
    primaryClass = "hep-ph",
    doi = "10.1016/j.physletb.2021.136494",
    journal = "Phys. Lett. B",
    volume = "820",
    pages = "136494",
    year = "2021"
}

@article{Xu:2025cyj,
    author = "Xu, Z. -N. and Yao, Z. -Q. and Binosi, D. and Ding, M. and Roberts, C. D. and Rodr{\'\i}guez-Quintero, J.",
    title = "{Distribution functions of a radially excited pion}",
    archivePrefix = "arXiv",
    primaryClass = "hep-ph",
    reportNumber = "NJU-INP 096/25",
    journal = "Eur. Phys. J. C",
    volume = "85",
    number = "3",
    pages = "330",
    year = "2025"
}

@article{Xu:2024nzp,
    author = "Xu, Zhen-Ni and Binosi, Daniele and Chen, Chen and Raya, Kh\'epani and Roberts, Craig D. and Rodr\'\i{}guez-Quintero, Jos\'e",
    title = "{Kaon distribution functions from empirical information}",
    archivePrefix = "arXiv",
    primaryClass = "hep-ph",
    reportNumber = "NJU-INP 094/24, USTC-ICTS/PCFT-24-48",
    journal = "Phys. Lett. B",
    volume = "865",
    pages = "139451",
    year = "2025"
}

@article{Raya:2024glv,
    author = "Raya, Kh{\'e}pani and Bashir, Adnan and Rodr{\'\i}guez-Quintero, Jos{\'e}",
    title = "{Mapping Spatial Distributions within Pseudoscalar Mesons}",
    archivePrefix = "arXiv",
    primaryClass = "hep-ph",
    journal = "Chin. Phys. Lett.",
    volume = "42",
    number = "2",
    pages = "020201",
    year = "2025"
}

@article{Chavez:2021llq,
    author = "Chavez, Jos\'e Manuel Morgado and Bertone, Valerio and De Soto Borrero, Feliciano and Defurne, Maxime and Mezrag, C\'edric and Moutarde, Herv\'e and Rodr\'\i{}guez-Quintero, Jos\'e and Segovia, Jorge",
    title = "{Pion generalized parton distributions: A path toward phenomenology}",
    archivePrefix = "arXiv",
    primaryClass = "hep-ph",
    journal = "Phys. Rev. D",
    volume = "105",
    number = "9",
    pages = "094012",
    year = "2022"
}

@article{Pasquini:2014ppa,
    author = "Pasquini, B. and Schweitzer, P.",
    title = "{Pion transverse momentum dependent parton distributions in a light-front constituent approach, and the Boer-Mulders effect in the pion-induced Drell-Yan process}",
    archivePrefix = "arXiv",
    primaryClass = "hep-ph",
    journal = "Phys. Rev. D",
    volume = "90",
    number = "1",
    pages = "014050",
    year = "2014"
}

@article{Raya:2024ejx,
    author = "Raya, Kh\'epani and Bashir, Adnan and Binosi, Daniele and Roberts, Craig D. and Rodr\'\i{}guez-Quintero, Jos\'e",
    title = "{Pseudoscalar Mesons and Emergent Mass}",
    archivePrefix = "arXiv",
    primaryClass = "hep-ph",
    reportNumber = "NJU-INP 084/24",
    journal = "Few Body Syst.",
    volume = "65",
    number = "2",
    pages = "60",
    year = "2024"
}

@article{Epelbaum:2022fjc,
    author = "Epelbaum, E. and Gegelia, J. and Lange, N. and Mei\ss{}ner, U. -G. and Polyakov, M. V.",
    title = "{Definition of Local Spatial Densities in Hadrons}",
    archivePrefix = "arXiv",
    journal = "Phys. Rev. Lett.",
    volume = "129",
    number = "1",
    pages = "012001",
    year = "2022"
}

@article{Freese:2021mzg,
    author = "Freese, Adam and Miller, Gerald A.",
    title = "{Unified formalism for electromagnetic and gravitational probes: Densities}",
    archivePrefix = "arXiv",
    reportNumber = "NT@UW-21-09",
    journal = "Phys. Rev. D",
    volume = "105",
    number = "1",
    pages = "014003",
    year = "2022"
}

@article{Jaffe:2020ebz,
    author = "Jaffe, Robert L.",
    title = "{Ambiguities in the definition of local spatial densities in light hadrons}",
    archivePrefix = "arXiv",
    reportNumber = "MIT-CTP/5253",
    journal = "Phys. Rev. D",
    volume = "103",
    number = "1",
    pages = "016017",
    year = "2021"
}

@article{Xing:2025eip,
    author = "Xing, Hui-Yu and Bian, Wen-Hao and Cui, Zhu-Fang and Roberts, Craig D.",
    title = "{Kaon and Pion Fragmentation Functions -- arXiv:2504.08142 [hep-ph]}",
    archivePrefix = "arXiv",
    primaryClass = "hep-ph",
    reportNumber = "NJU-INP 099/25",
    journal = "Eur. Phys. J. C",
    month = "4",
    pages = "\emph{in press}",
    year = "2025"
}

@article{Mezrag:2023nkp,
    author = "Mezrag, C\'edric",
    title = "{Generalised Parton Distributions in Continuum Schwinger Methods: Progresses, Opportunities and Challenges}",
    journal = "Particles",
    volume = "6",
    number = "1",
    pages = "262--296",
    year = "2023"
}

@article{Accardi:2023chb,
    author = "Accardi, A. and others",
    title = "{Strong Interaction Physics at the Luminosity Frontier with 22 GeV Electrons at Jefferson Lab
                 -- arXiv:2306.09360 [nucl-ex]}",
    archivePrefix = "arXiv",
    reportNumber = "JLAB-THY-23-3848",
    month = "6",
    year = "2023"
}

@article{Yin:2023dbw,
    author = "Yin, Pei-Lin and Xu, Yin-Zhen and Cui, Zhu-Fang and Roberts, Craig D. and Rodr\'\i{}guez-Quintero, Jos\'e",
    title = "{All-Orders Evolution of Parton Distributions: Principle, Practice, and Predictions}",
    archivePrefix = "arXiv",
    reportNumber = "NJU-INP 075/23",
    journal = "Chin. Phys. Lett. \emph{Express}",
    volume = "40",
    number = {9},
    pages = {091201},
    year = "2023"
}

@article{Xu:2023bwv,
    author = "Xu, Yin-Zhen and Raya, Kh\'epani and Cui, Zhu-Fang and Roberts, Craig D. and Rodr\'\i{}guez-Quintero, J.",
    title = "{Empirical Determination of the Pion Mass Distribution}",
    archivePrefix = "arXiv",
    reportNumber = "NJU-INP 070/23",
    journal = "Chin. Phys. Lett. \emph{Express}",
    volume = "40",
    number = "4",
    pages = "041201",
    year = "2023"
}

@article{Chavez:2021koz,
    author = "Ch\'avez, Jos\'e Manuel Morgado and Bertone, Valerio and De Soto Borrero, Feliciano and Defurne, Maxime and Mezrag, C\'edric and Moutarde, Herv\'e and Rodr\'\i{}guez-Quintero, Jos\'e and Segovia, Jorge",
    title = "{Accessing the Pion 3D Structure at US and China Electron-Ion Colliders}",
    archivePrefix = "arXiv",
    primaryClass = "hep-ph",
    journal = "Phys. Rev. Lett.",
    volume = "128",
    number = "20",
    pages = "202501",
    year = "2022"
}

@article{Mezrag:2022pqk,
    author = "Mezrag, C\'edric",
    title = "{An Introductory Lecture on Generalised Parton Distributions}",
    archivePrefix = "arXiv",
    primaryClass = "hep-ph",
    journal = "Few Body Syst.",
    volume = "63",
    number = "3",
    pages = "62",
    year = "2022"
}

@article{Albino:2022gzs,
    author = "Albino, L. and Higuera-Angulo, I. M. and Raya, K. and Bashir, A.",
    title = "{Pseudoscalar mesons: Light front wave functions, GPDs, and PDFs}",
    archivePrefix = "arXiv",
    primaryClass = "hep-ph",
    journal = "Phys. Rev. D",
    volume = "106",
    number = "3",
    pages = "034003",
    year = "2022"
}

@article{Quintans:2022utc,
    author = "Quintans, Catarina",
    title = "{The New AMBER Experiment at the CERN SPS}",
    journal = "Few Body Syst.",
    volume = "63",
    number = "4",
    pages = "72",
    year = "2022"
}

@article{Lu:2022cjx,
    author = "Lu, Ya and Chang, Lei and Raya, Kh\'epani and Roberts, Craig D. and Rodr\'\i{}guez-Quintero, Jos\'e",
    title = "{Proton and pion distribution functions in counterpoint}",
    archivePrefix = "arXiv",
    reportNumber = "NJU-INP 056/22",
    journal = "Phys. Lett. B",
    volume = "830",
    pages = "137130",
    year = "2022"
}

@article{Cui:2022bxn,
    author = "Cui, Z. -F. and Ding, Minghui and Morgado, J. M. and Raya, K. and Binosi, D. and Chang, L. and De Soto, F. and Roberts, C. D. and Rodr\'\i{}guez-Quintero, J. and Schmidt, S. M.",
    title = "{Emergence of pion parton distributions}",
    archivePrefix = "arXiv",
    reportNumber = "NJU-INP 054/22",
    journal = "Phys. Rev. D",
    volume = "105",
    number = "9",
    pages = "L091502",
    year = "2022"
}

@article{Cui:2021mom,
    author = "Cui, Z. -F. and Ding, Minghui and Morgado, J. M. and Raya, K. and Binosi, D. and Chang, L. and Papavassiliou, J. and Roberts, C. D. and Rodr\'\i{}guez-Quintero, J. and Schmidt, S. M.",
    title = "{Concerning pion parton distributions}",
    archivePrefix = "arXiv",
    reportNumber = "NJU-INP 053/21",
    journal = "Eur. Phys. J. A",
    volume = "58",
    number = "1",
    pages = "10",
    year = "2022"
}

@article{Raya:2021zrz,
    author = "Raya, Kh{\'e}pani and Cui, Zhu-Fang and Chang, Lei and Morgado, Jos{\'e}-Manuel and
                Roberts, Craig D. and Rodr{\'{\i}}guez-Quintero, Jos{\'e}",
    title = "{Revealing pion and kaon structure via generalised parton distributions}",
    archivePrefix = "arXiv",
    primaryClass = "hep-ph",
    reportNumber = "NJU-INP 051/21",
    journal = "Chin. Phys. C",
    volume = "46",
    number = "26",
    pages = "013105",
    year = "2022"
}

@article{Brodsky:1979gy,
    author = "Brodsky, Stanley J. and Lepage, G. Peter",
    title = "{Perturbative Quantum Chromodynamics}",
    reportNumber = "SLAC-PUB-2447",
    journal = "Prog. Math. Phys.",
    volume = "4",
    pages = "255--422",
    month = "12",
    year = "1979"
}

@article{Zhang:2021mtn,
    author = "Zhang, Jin-Li and Raya, Kh\'epani and Chang, Lei and Cui, Zhu-Fang and Morgado, Jos\'e Manuel and Roberts, Craig D and Rodr\'\i{}guez-Quintero, Jose",
    title = "{Measures of pion and kaon structure from generalised parton distributions}",
    archivePrefix = "arXiv",
    primaryClass = "hep-ph",
    reportNumber = "NJU-INP 032/21",
    journal = "Phys. Lett. B",
    volume = "815",
    pages = "136158",
    year = "2021"
}

@article{Arrington:2021biu,
    author = "Arrington, J. and others",
    title = "{Revealing the structure of light pseudoscalar mesons at the electron\textendash{}ion collider}",
    archivePrefix = "arXiv",
    journal = "J. Phys. G",
    volume = "48",
    pages = "075106",
    year = "2021"
}

@article{Roberts:2021nhw,
    author = "Roberts, Craig D. and Richards, David G. and Horn, Tanja and Chang, Lei",
    title = "{Insights into the emergence of mass from studies of pion and kaon structure}",
    archivePrefix = "arXiv",
    reportNumber = "NJU-INP 034/21",
    journal = "Prog. Part. Nucl. Phys.",
    volume = "120",
    pages = "103883",
    year = "2021"
}

@article{Belitsky:2005qn,
    author = "Belitsky, A.V. and Radyushkin, A.V.",
    title = "{Unraveling hadron structure with generalized parton distributions}",
    archivePrefix = "arXiv",
    reportNumber = "JLAB-THY-04-34",
    journal = "Phys. Rept.",
    volume = "418",
    pages = "1--387",
    year = "2005"
}

@article{Polyakov:2018zvc,
    author = "Polyakov, Maxim V. and Schweitzer, Peter",
    title = "{Forces inside hadrons: pressure, surface tension, mechanical radius, and all that}",
    archivePrefix = "arXiv",
    primaryClass = "hep-ph",
    journal = "Int. J. Mod. Phys. A",
    volume = "33",
    number = "26",
    pages = "1830025",
    year = "2018"
}

@article{Qin:2020rad,
    author = "Qin, Si-Xue and Roberts, Craig D",
    title = "{Impressions of the Continuum Bound State Problem in QCD}",
    archivePrefix = "arXiv",
    primaryClass = "hep-ph",
    reportNumber = "NJU-INP 023/20",
    journal = "Chin. Phys. Lett.",
    volume = "37",
    pages = "121201",
    number = "12",
    year = "2020"
}

@article{Cui:2020tdf,
    author = "Cui, Zhu-Fang and Ding, Minghui and Gao, Fei and Raya, Kh\'epani and Binosi, Daniele
                    and Chang, Lei and Roberts, Craig D and Rodr\'{\i}guez-Quintero, Jose and Schmidt, Sebastian M",
    title = "{Kaon and pion parton distributions}",
    journal = "Eur. Phys. J. C",
    volume = "80",
    pages = "1064",
    year = "2020"
}

@article{Cui:2020dlm,
    author = "Cui, Zhu-Fang and Ding, Minghui and Gao, Fei and Raya, Kh\'epani and Binosi, Daniele
            and Chang, Lei and Roberts, Craig D and Rodr\'{\i}guez-Quintero, Jose and Schmidt, Sebastian M",
    title = "{Higgs modulation of emergent mass as revealed in kaon and pion parton distributions}",
    archivePrefix = "arXiv",
    primaryClass = "hep-ph",
    reportNumber = "NJU-INP 019/20",
    journal = "Eur. Phys. J. A (Lett.)",
    volume = "57",
    number = "1",
    pages = "5",
    year = "2021"
}

@article{Anderle:2021wcy,
    author = "Anderle, Daniele P. and others",
    title = "{Electron-ion collider in China}",
    archivePrefix = "arXiv",
    journal = "Front. Phys. (Beijing)",
    volume = "16",
    number = "6",
    pages = "64701",
    year = "2021"
}

@article{Ding:2019qlr,
    author = "Ding, Minghui and Raya, Kh\'epani and Binosi, Daniele and Chang, Lei and Roberts,
                Craig D and Schmidt, Sebastian M",
      title          = "{Drawing insights from pion parton distributions}",
      journal        = "Chin. Phys. C (Lett.)",
      volume         = "44",
      year           = "2020",
      pages          = "031002",
      archivePrefix  = "arXiv",
      primaryClass   = "hep-ph",
      reportNumber   = "NJU-INP 012/19",
      SLACcitation   = "%%CITATION = ARXIV:1912.07529;%%"
}

@article{Ding:2019lwe,
    author = "Ding, Minghui and Raya, Kh\'epani and Binosi, Daniele and Chang, Lei
                and Roberts, Craig D and Schmidt, Sebastian M.",
    title = "{Symmetry, symmetry breaking, and pion parton distributions}",
    archivePrefix = "arXiv",
    primaryClass = "nucl-th",
    reportNumber = "NJU-INP 003/19",
    journal = "Phys. Rev. D",
    volume = "101",
    number = "5",
    pages = "054014",
    year = "2020"
}

@article{Chouika:2017dhe,
      author         = "Chouika, N. and Mezrag, C. and Moutarde, H. and
                        Rodr{\'{\i}}guez-Quintero, J.",
      title          = "{Covariant Extension of the GPD overlap representation at
                        low Fock states}",
      journal        = "Eur. Phys. J. C",
      volume         = "77",
      year           = "2017",
      pages          = "906",
      archivePrefix  = "arXiv",
      primaryClass   = "hep-ph",
      SLACcitation   = "%%CITATION = ARXIV:1711.05108;%%"
}

@article{Xu:2018eii,
      author         = "Xu, Shu-Sheng and Chang, Lei and Roberts, Craig D. and
                        Zong, Hong-Shi",
      title          = "{Pion and kaon valence-quark parton quasidistributions}",
      journal        = "Phys. Rev. D",
      volume         = "97",
      year           = "2018",
      pages          = "094014",
      archivePrefix  = "arXiv",
      SLACcitation   = "%%CITATION = ARXIV:1802.09552;%%"
}

@article{Chouika:2017rzs,
      author         = "Chouika, N. and Mezrag, C. and Moutarde, H. and
                        Rodr{\'{\i}}guez-Quintero, J.",
      title          = "{A Nakanishi-based model illustrating the covariant
                        extension of the pion GPD overlap representation and its
                        ambiguities}",
      journal        = "Phys. Lett. B",
      volume         = "780",
      year           = "2018",
      pages          = "287-293",
      archivePrefix  = "arXiv",
      primaryClass   = "hep-ph",
      SLACcitation   = "%%CITATION = ARXIV:1711.11548;%%"
}

@article{Mezrag:2016hnp,
      author         = "Mezrag, C. and Moutarde, H. and Rodr{\'i}guez-Quintero, J.",
      title          = "{From Bethe-Salpeter Wave functions to Generalised
                        Parton Distributions}",
      journal        = "Few Body Syst.",
      volume         = "57",
      year           = "2016",
      pages          = "729-772",
      archivePrefix  = "arXiv",
      primaryClass   = "nucl-th",
      SLACcitation   = "%%CITATION = ARXIV:1602.07722;%%"
}

@article{Eichmann:2016yit,
      author         = "Eichmann, Gernot and Sanchis-Alepuz, Helios and Williams,
                        Richard and Alkofer, Reinhard and Fischer, Christian S.",
      title          = "{Baryons as relativistic three-quark bound states}",
      journal        = "Prog. Part. Nucl. Phys.",
      volume         = "91",
      year           = "2016",
      pages          = "1-100",
      archivePrefix  = "arXiv",
      SLACcitation   = "%%CITATION = ARXIV:1606.09602;%%"
}

@article{Roberts:2016vyn,
      author         = "Roberts, Craig D.",
      title          = "{Perspective on the origin of hadron masses}",
      journal        = "Few Body Syst.",
      volume         = "58",
      year           = "2017",
      pages          = "5",
      archivePrefix  = "arXiv",
      primaryClass   = "nucl-th",
      SLACcitation   = "%%CITATION = ARXIV:1606.03909;%%"
}

@article{Chen:2016sno,
      author         = "Chen, Chen and Chang, Lei and Roberts, Craig D. and Wan,
                        Shaolong and Zong, Hong-Shi",
      title          = "{Valence-quark distribution functions in the kaon and
                        pion}",
  journal = {Phys. Rev. D},
  volume = {93},
  pages = {074021},
      year           = "2016",
      archivePrefix  = "arXiv",
      primaryClass   = "nucl-th",
      SLACcitation   = "%%CITATION = ARXIV:1602.01502;%%"
}

@article{Horn:2016rip,
      author         = "Horn, Tanja and Roberts, Craig D.",
      title          = "{The pion: an enigma within the Standard Model}",
      journal        = "J. Phys. G.",
      volume         = "43",
      year           = "2016",
      pages          = "073001",
      archivePrefix  = "arXiv",
      SLACcitation   = "%%CITATION = ARXIV:1602.04016;%%"
}

@article{Miller:2010tz,
      author         = "Miller, G. A. and Strikman, M. and Weiss, C.",
      title          = "{Pion transverse charge density from timelike form factor
                        data}",
      journal        = "Phys. Rev. D",
      volume         = "83",
      year           = "2011",
      pages          = "013006",
      doi            = "10.1103/PhysRevD.83.013006",
      archivePrefix  = "arXiv",
      primaryClass   = "hep-ph",
      reportNumber   = "NT@UW-10-20, JLAB-THY-10-1274",
      SLACcitation   = "%%CITATION = ARXIV:1011.1472;%%"
}

@article{Binosi:2014aea,
      author         = "Binosi, Daniele and Chang, Lei and Papavassiliou, Joannis
                        and Roberts, Craig D.",
      title          = "{Bridging a gap between continuum-QCD and \emph{ab initio}
                        predictions of hadron observables}",
      journal        = "Phys. Lett. B",
      volume         = "742",
      pages          = "183-188",
      year           = "2015",
      archivePrefix  = "arXiv",
      primaryClass   = "nucl-th",
      reportNumber   = "ADP-14-42-T901",
      SLACcitation   = "%%CITATION = ARXIV:1412.4782;%%"
}

@article{Accardi:2012qut,
      author         = "Accardi, A. and others",
      editor         = "Deshpande, A. and Meziani, Z. E. and Qiu, J. W.",
      title          = "{Electron Ion Collider: The Next QCD Frontier}",
      journal        = "Eur. Phys. J. A",
      volume         = "52",
      year           = "2016",
      pages          = "268",
      archivePrefix  = "arXiv",
      primaryClass   = "nucl-ex",
      reportNumber   = "BNL-98815-2012-JA, JLAB-PHY-12-1652",
      SLACcitation   = "%%CITATION = ARXIV:1212.1701;%%"
}

@article{Polyakov:1998ze,
      author         = "Polyakov, Maxim V.",
      title          = "{Hard exclusive electroproduction of two pions and their
                        resonances}",
      journal        = "Nucl. Phys. B",
      volume         = "555",
      pages          = "231",
      doi            = "10.1016/S0550-3213(99)00314-4",
      year           = "1999",
      archivePrefix  = "arXiv",
      primaryClass   = "hep-ph",
      reportNumber   = "RUB-TPII-14-98",
      SLACcitation   = "%%CITATION = HEP-PH/9809483;%%",
}

@article{Burkardt:2000za,
      author         = "Burkardt, Matthias",
      title          = "{Impact parameter dependent parton distributions and off
                        forward parton distributions for {$\zeta > 0$}}",
      journal        = "Phys. Rev. D",
      volume         = "62",
      pages          = "071503",
      year           = "2000",
      archivePrefix  = "arXiv",
      primaryClass   = "hep-ph",
      SLACcitation   = "%%CITATION = HEP-PH/0005108;%%"
}

@article{Mezrag:2014jka,
      author         = "Mezrag, C. and Chang, L. and Moutarde, H. and Roberts,
                        C. D. and Rodr{\'i}guez-Quintero, J. and Sabati{\'e}, F. and Schmidt, S. M.",
      title          = "{Sketching the pion's valence-quark generalised parton
                        distribution}",
      journal        = "Phys. Lett. B",
      volume         = "741",
      pages          = "190-196",
      year           = "2015",
      archivePrefix  = "arXiv",
      primaryClass   = "nucl-th",
      SLACcitation   = "%%CITATION = ARXIV:1411.6634;%%",
}

@article{Burkardt:2002hr,
      author         = "Burkardt, Matthias",
      title          = "{Impact parameter space interpretation for generalized
                        parton distributions}",
      journal        = "Int. J. Mod. Phys. A",
      volume         = "18",
      pages          = "173-208",
      year           = "2003",
      archivePrefix  = "arXiv",
      primaryClass   = "hep-ph",
      SLACcitation   = "%%CITATION = HEP-PH/0207047;%%"
}

@article{Ji:1996nm,
      author         = "Ji, Xiang-Dong",
      title          = "{Deeply virtual Compton scattering}",
      journal        = "Phys. Rev. D",
      volume         = "55",
      pages          = "7114-7125",
      doi            = "10.1103/PhysRevD.55.7114",
      year           = "1997",
      archivePrefix  = "arXiv",
      primaryClass   = "hep-ph",
      reportNumber   = "UMD-PP-97-26, MIT-CTP-2568",
      SLACcitation   = "%%CITATION = HEP-PH/9609381;%%"
}

@article{Radyushkin:1996nd,
      author         = "Radyushkin, A.V.",
      title          = "{Scaling limit of deeply virtual Compton scattering}",
      journal        = "Phys. Lett. B",
      volume         = "380",
      pages          = "417-425",
      year           = "1996",
      archivePrefix  = "arXiv",
      primaryClass   = "hep-ph",
      reportNumber   = "CEBAF-TH-96-05",
      SLACcitation   = "%%CITATION = HEP-PH/9604317;%%"
}

@article{Chang:2014lva,
      author         = "Chang, Lei and Mezrag, C{\'e}dric and Moutarde, Herv{\'e} and
                        Roberts, Craig D. and Rodr{\'{\i}}guez-Quintero, Jose and
                        Tandy, Peter C.",
      title          = "{Basic features of the pion valence-quark distribution
                        function}",
      journal        = "Phys. Lett. B",
      volume         = "737",
      pages          = "23-29",
      year           = "2014",
      archivePrefix  = "arXiv",
      primaryClass   = "nucl-th",
      reportNumber   = "ADP-14-20-T878",
      SLACcitation   = "%%CITATION = ARXIV:1406.5450;%%"
}

@article{Diehl:2003ny,
      author         = "Diehl, M.",
      title          = "{Generalized parton distributions}",
      journal        = "Phys. Rept.",
      volume         = "388",
      pages          = "41-277",
      year           = "2003",
      archivePrefix  = "arXiv",
      primaryClass   = "hep-ph",
      reportNumber   = "DESY-THESIS-2003-018",
      SLACcitation   = "%%CITATION = HEP-PH/0307382;%%"
}

@Article{Dittes:1988xz,
     author    = "Dittes, F. M. and M{\"u}ller, Dieter and Robaschik, D. and
                  Geyer, B. and Ho{\v{r}}ej{\v{s}}i, J.",
     title     = "{The Altarelli-Parisi Kernel as Asymptotic Limit of an
                  Extended Brodsky-Lepage Kernel}",
     journal   = "Phys. Lett. B",
     volume    = "209",
     year      = "1988",
     pages     = "325-329",
     doi       = "10.1016/0370-2693(88)90955-0",
     SLACcitation  = "%%CITATION = PHLTA,B209,325;%%"
}

@article{Cloet:2013tta,
      author         = "Cloet, Ian C. and Chang, Lei and Roberts, Craig D.
                        and Schmidt, Sebastian M. and Tandy, Peter C.",
      title          = "{Pion distribution amplitude from lattice-QCD}",
      journal        = "Phys. Rev. Lett.",
      volume         = "111",
      pages          = "092001",
      year           = "2013",
      archivePrefix  = "arXiv",
      primaryClass   = "nucl-th",
      SLACcitation   = "%%CITATION = ARXIV:1306.2645;%%"
}

@article{Sachs:1962zzc,
      author         = "Sachs, R.G.",
      title          = "{High-Energy Behavior of Nucleon Electromagnetic Form
                        Factors}",
      journal        = "Phys. Rev.",
      volume         = "126",
      pages          = "2256-2260",
      year           = "1962",
      SLACcitation   = "%%CITATION = PHRVA,126,2256;%%"
}

@article{Chang:2013pq,
      author         = "Chang, Lei and Cloet, I. C. and Cobos-Martinez, J. J. and
                        Roberts, C. D. and Schmidt, S. M. and Tandy, Peter C.",
      title          = "{Imaging dynamical chiral symmetry breaking: pion wave
                        function on the light front}",
      journal        = "Phys. Rev. Lett.",
      volume    = "110",
      pages     = "132001",
      year           = "2013",
      archivePrefix  = "arXiv",
      SLACcitation   = "%%CITATION = ARXIV:1301.0324;%%"
}

@article{Roberts:2012svX,
      author         = "Roberts, Craig D.",
      title          = "{Strong QCD and Dyson-Schwinger Equations}",
      note           = "{arXiv:1203.5341 [nucl-th]}",
      journal        = "IRMA Lect. Math. and Theor. Phys.",
      volume         = "21",
      pages          = "356-458",
      year           = "2015",
      archivePrefix  = "arXiv",
      primaryClass   = "nucl-th",
      SLACcitation   = "%%CITATION = ARXIV:1203.5341;%%"
}

@Article{Lepage:1980fj,
     author    = "Lepage, G. Peter and Brodsky, Stanley J.",
     title     = "{Exclusive Processes in Perturbative Quantum
                  Chromodynamics}",
     journal   = "Phys. Rev. D",
     volume    = "22",
     year      = "1980",
     pages     = "2157-2198",
     SLACcitation  = "%%CITATION = PHRVA,D22,2157;%%"
}

@article{Yao:2025xjx,
    author = "Yao, Zhao-Qian and Xu, Zhen-Ni and Xiao, Yu-Yang and Roberts, Craig D. and Rodriguez-Quintero, Jose",
    title = "{Symmetry-preserving calculation of pion light-front wave functions}",
    eprint = "2512.13938",
    archivePrefix = "arXiv",
    primaryClass = "hep-ph",
    reportNumber = "NJU-INP 110-25",
    month = "12",
    year = "2025"
}

@article{Maris:1997tm,
    author = "Maris, Pieter and Roberts, Craig D.",
    title = "{Pi- and K meson Bethe-Salpeter amplitudes}",
    eprint = "nucl-th/9708029",
    archivePrefix = "arXiv",
    reportNumber = "ANL-PHY-8788-TH-97",
    doi = "10.1103/PhysRevC.56.3369",
    journal = "Phys. Rev. C",
    volume = "56",
    pages = "3369--3383",
    year = "1997"
}

@article{Qin:2011dd,
    author = "Qin, Si-xue and Chang, Lei and Liu, Yu-xin and Roberts, Craig D. and Wilson, David J.",
    title = "{Interaction model for the gap equation}",
    eprint = "1108.0603",
    archivePrefix = "arXiv",
    primaryClass = "nucl-th",
    doi = "10.1103/PhysRevC.84.042202",
    journal = "Phys. Rev. C",
    volume = "84",
    pages = "042202",
    year = "2011"
}

@article{Raya:2026gwg,
    author = "Raya, Kh{\'e}pani and Xu, Zhen-Ni and Yao, Zhao-Qian and Rodr{\'\i}guez-Quintero, Jos{\'e}",
    title = "{Pion structure from its light-front wave function}",
    eprint = "2606.20998",
    archivePrefix = "arXiv",
    primaryClass = "hep-ph",
    month = "6",
    year = "2026"
}

@article{Xu:2024vkn,
    author = "Xu, Y. -Z. and Raya, K. and Rodr{\'\i}guez-Quintero, J. and Segovia, J.",
    title = "{Charge distributions of pseudoscalar and vector mesons from Dyson-Schwinger equations}",
    eprint = "2406.13306",
    archivePrefix = "arXiv",
    primaryClass = "hep-ph",
    doi = "10.1103/PhysRevD.110.054031",
    journal = "Phys. Rev. D",
    volume = "110",
    number = "5",
    pages = "054031",
    year = "2024"
}

@article{Almeida-Zamora:2023bqb,
    author = "Almeida-Zamora, B. and Cobos-Mart{\'\i}nez, J. J. and Bashir, A. and Raya, K. and Rodr{\'\i}guez-Quintero, J. and Segovia, J.",
    title = "{Algebraic model to study the internal structure of pseudoscalar mesons with heavy-light quark content}",
    eprint = "2309.17282",
    archivePrefix = "arXiv",
    primaryClass = "hep-ph",
    reportNumber = "JLAB-THY-23-3977",
    doi = "10.1103/PhysRevD.109.014016",
    journal = "Phys. Rev. D",
    volume = "109",
    number = "1",
    pages = "014016",
    year = "2024"
}

@article{Burkardt:2015qoa,
    author = "Burkardt, M. and Pasquini, B.",
    title = "{Modelling the nucleon structure}",
    eprint = "1510.02567",
    archivePrefix = "arXiv",
    primaryClass = "hep-ph",
    doi = "10.1140/epja/i2016-16161-7",
    journal = "Eur. Phys. J. A",
    volume = "52",
    number = "6",
    pages = "161",
    year = "2016"
}

@article{Xing:2023eed,
    author = "Xing, Zanbin and Ding, Minghui and Raya, Kh{\'e}pani and Chang, Lei",
    title = "{Fresh look at the generalized parton distributions of light pseudoscalar mesons}",
    eprint = "2301.02958",
    archivePrefix = "arXiv",
    primaryClass = "hep-ph",
    doi = "10.1140/epja/s10050-024-01256-z",
    journal = "Eur. Phys. J. A",
    volume = "60",
    number = "2",
    pages = "33",
    year = "2024"
}

@article{Zeng:2026peb,
    author = "Zeng, X. -Y. and Xiao, Y. -Y. and Xu, Z. -N. and Roberts, C. D. and Rodr{\'\i}guez-Quintero, J.",
    title = "{Distribution amplitudes and functions of ground-state scalar and pseudoscalar charmonia}",
    eprint = "2604.06510",
    archivePrefix = "arXiv",
    primaryClass = "hep-ph",
    reportNumber = "NJU-INP 118/26",
    month = "4",
    year = "2026"
}

@article{Muller:1994ses,
    author = {M{\"u}ller, Dieter and Robaschik, D. and Geyer, B. and Dittes, F. -M. and Ho{\v{r}}ej{\v{s}}i, J.},
    title = "{Wave functions, evolution equations and evolution kernels from light ray operators of QCD}",
    eprint = "hep-ph/9812448",
    archivePrefix = "arXiv",
    reportNumber = "NTZ-6-91, NTZ-91-6",
    doi = "10.1002/prop.2190420202",
    journal = "Fortsch. Phys.",
    volume = "42",
    pages = "101--141",
    year = "1994"
}

@article{Raya:2022eqa,
    author = "Raya, Kh{\'e}pani and Rodr{\'\i}guez-Quintero, Jos{\'e}",
    title = "{Highlights of pion and kaon structure from continuum analyses}",
    eprint = "2204.01642",
    archivePrefix = "arXiv",
    primaryClass = "hep-ph",
    doi = "10.31349/SuplRevMexFis.3.0308008",
    journal = "Rev. Mex. Fis. Suppl.",
    volume = "3",
    number = "3",
    pages = "0308008",
    year = "2022"
}

@article{Almeida-Zamora:2023rwg,
    author = "Almeida-Zamora, B. and Cobos-Mart{\'\i}nez, J. J. and Bashir, A. and Raya, K. and Rodr{\'\i}guez-Quintero, J. and Segovia, J.",
    title = "{Light-front wave functions of vector mesons in an algebraic model}",
    eprint = "2303.09581",
    archivePrefix = "arXiv",
    primaryClass = "hep-ph",
    reportNumber = "JLAB-THY-23-3779",
    doi = "10.1103/PhysRevD.107.074037",
    journal = "Phys. Rev. D",
    volume = "107",
    number = "7",
    pages = "074037",
    year = "2023"
}

@article{Cheng:2026nud,
    author = "Cheng, Dan-Dan and Ding, Minghui and Binosi, Daniele and Roberts, Craig D.",
    title = "{Kaon Boer-Mulders function using a contact interaction}",
    eprint = "2603.25941",
    archivePrefix = "arXiv",
    primaryClass = "hep-ph",
    reportNumber = "NJU-INP 117/26",
    month = "3",
    year = "2026"
}

@article{Hofstadter:1958psf,
    author = "Hofstadter, R. and Bumiller, F. and Yearian, M. R.",
    title = "{Electromagnetic Structure of the Proton and Neutron}",
    doi = "10.1103/RevModPhys.30.482",
    journal = "Rev. Mod. Phys.",
    volume = "30",
    number = "2",
    pages = "482",
    year = "1958"
}

@article{Ernst:1960zza,
    author = "Ernst, F. J. and Sachs, R. G. and Wali, K. C.",
    title = "{Electromagnetic form factors of the nucleon}",
    doi = "10.1103/PhysRev.119.1105",
    journal = "Phys. Rev.",
    volume = "119",
    pages = "1105--1114",
    year = "1960"
}

@article{Miller:2018ybm,
    author = "Miller, Gerald A.",
    title = "{Defining the proton radius: A unified treatment}",
    eprint = "1812.02714",
    archivePrefix = "arXiv",
    primaryClass = "nucl-th",
    reportNumber = "NT@UW-18-20",
    doi = "10.1103/PhysRevC.99.035202",
    journal = "Phys. Rev. C",
    volume = "99",
    number = "3",
    pages = "035202",
    year = "2019"
}

@article{Chen:2022smg,
    author = "Chen, Yi and Lorc{\'e}, C{\'e}dric",
    title = "{Pion and nucleon relativistic electromagnetic four-current distributions}",
    eprint = "2210.02908",
    archivePrefix = "arXiv",
    primaryClass = "hep-ph",
    doi = "10.1103/PhysRevD.106.116024",
    journal = "Phys. Rev. D",
    volume = "106",
    number = "11",
    pages = "116024",
    year = "2022"
}

@article{Lorce:2022jyi,
    author = "Lorc{\'e}, C{\'e}dric and Wang, Pierre",
    title = "{Deuteron relativistic charge distributions}",
    eprint = "2204.01465",
    archivePrefix = "arXiv",
    primaryClass = "hep-ph",
    doi = "10.1103/PhysRevD.105.096032",
    journal = "Phys. Rev. D",
    volume = "105",
    number = "9",
    pages = "096032",
    year = "2022"
}

@article{Lorce:2020onh,
    author = "Lorc{\'e}, C{\'e}dric",
    title = "{Charge Distributions of Moving Nucleons}",
    eprint = "2007.05318",
    archivePrefix = "arXiv",
    primaryClass = "hep-ph",
    doi = "10.1103/PhysRevLett.125.232002",
    journal = "Phys. Rev. Lett.",
    volume = "125",
    number = "23",
    pages = "232002",
    year = "2020"
}

@article{Chen:2023dxp,
    author = "Chen, Yi and Lorc{\'e}, C{\'e}dric",
    title = "{Nucleon relativistic polarization and magnetization distributions}",
    eprint = "2302.04672",
    archivePrefix = "arXiv",
    primaryClass = "hep-ph",
    doi = "10.1103/PhysRevD.107.096003",
    journal = "Phys. Rev. D",
    volume = "107",
    number = "9",
    pages = "096003",
    year = "2023"
}

@article{Miramontes:2025vzb,
    author = "Miramontes, A. S. and Papavassiliou, J. and Pawlowski, J. M.",
    title = "{Electromagnetic properties of heavy-light mesons}",
    eprint = "2508.20631",
    archivePrefix = "arXiv",
    primaryClass = "hep-ph",
    doi = "10.1140/epjc/s10052-025-15121-w",
    journal = "Eur. Phys. J. C",
    volume = "85",
    number = "12",
    pages = "1390",
    year = "2025"
}

\end{document}